\documentclass[fleqn,usenatbib]{mnras}

\usepackage{newtxtext,newtxmath}

\usepackage[T1]{fontenc}
\usepackage{ae,aecompl}

\usepackage{graphicx}	
\usepackage{amsmath}	
\usepackage{amssymb}	

\newcommand{\iso}[2]{\hbox{${}^{#1}{\rm #2}$}}
\newcommand{\Msun}{\ensuremath{{M}_{\odot}}}
\newcommand{\Lsun}{\ensuremath{{L}_{\odot}}}

\title[Yields of SMC-AGB models]{Heavy-element yields and abundances of Asymptotic Giant Branch models with a Small Magellanic Cloud metallicity}

\author[A. I. Karakas et al.]{Amanda I. Karakas,$^{1}$\thanks{E-mail: amanda.karakas@monash.edu}
Maria Lugaro,$^{2,1}$
Mar\'ilia Carlos,$^{1,3}$
Borb\'ala Cseh,$^{2}$
Devika Kamath,$^{4,5}$
\newauthor
and D. A. Garc\'{\i}a-Hern\'andez$^{6,7}$
\\
$^{1}$Monash Centre for Astrophysics, School of Physics and Astronomy, Monash University, VIC 3800, Australia\\
$^{2}$Konkoly Observatory, Research Centre for Astronomy and Earth Sciences, Hungarian Academy of Sciences, H-1121 Budapest, Hungary\\
$^{3}$Universidade de S\~ao Paulo, IAG, Departamento de Astronomia, Rua do Mat\~ao 1226, Cidade Universit\'aria, 05508-900 S\~ao Paulo, SP, Brazil\\
$^{4}$Department of Physics and Astronomy, Macquarie University, Sydney, NSW, Australia\\
$^{5}$Australian Astronomical Observatory, PO Box 915, North Ryde, NSW 1670, Australia\\
$^{6}$Instituto de Astrof\'{\i}sica de Canarias (IAC), E-38205 La Laguna, Tenerife, Spain\\
$^{7}$Departamento de Astrof\'{\i}sica, Universidad de La Laguna (ULL), E-38206 La Laguna, Spain\\
}

\date{Accepted XXX. Received YYY; in original form ZZZ}

\pubyear{2018}

\begin{document}
\label{firstpage}
\pagerange{\pageref{firstpage}--\pageref{lastpage}}
\maketitle

\begin{abstract}
We present new theoretical stellar yields and surface abundances for asymptotic giant branch (AGB) models with a metallicity appropriate for stars in the Small Magellanic Cloud (SMC, $Z= 0.0028$, [Fe/H] $\approx -0.7$). New evolutionary sequences and post-processing nucleosynthesis results are presented for initial masses between 1$\Msun$ and 7$\Msun$, where the 7$\Msun$ is a super-AGB star with an O-Ne core. Models above 1.15$\Msun$ become carbon rich during the AGB, and hot bottom burning begins in models $M \ge 3.75 \Msun$. We present stellar surface abundances as a function of thermal pulse number for elements between C to Bi and for a selection of isotopic ratios for elements up to Fe and Ni (e.g., \iso{12}C/\iso{13}C), which can be compared to observations. The integrated stellar yields are presented for each model in the grid for hydrogen, helium and all stable elements from C to Bi. We present evolutionary sequences of intermediate-mass models between 4--7$\Msun$ and nucleosynthesis results for three masses ($M=3.75, 5, 7\Msun$) including $s$-process elements for two widely used AGB mass-loss
prescriptions. We discuss our new models in the context of evolved AGB stars and post-AGB stars in the Small Magellanic Clouds, barium stars in our Galaxy, the composition of Galactic globular clusters including Mg isotopes with a similar metallicity to our models, and to pre-solar grains which may have an origin in metal-poor AGB stars.
\end{abstract}




\section{Introduction}

The Large and Small Magellanic Clouds (LMC and SMC, respectively) are two of the best studied satellite galaxies around the Milky Way Galaxy. They are both dwarf irregular galaxies with an average metallicity lower by about a factor of 2 and 5 respectively compared to the Milky Way thin disc, and show on-going star formation. Because of their proximity to the Milky Way Galaxy it is possible to perform detailed studies of the individual stars and the stellar populations within them \citep[e.g.,][]{cole05,mucciarelli06,muccia08,mackey08,milone09}. Both the LMC and SMC harbour populations of young massive stars 
\citep[e.g., the Tarantula nebula in the LMC,][]{evans11} as well as thousands of evolved asymptotic giant branch stars including many carbon stars \citep{wood83,frogel90,sloan08,melbourne13}. 

The population of asymptotic giant branch (AGB) in the Magellanic Clouds are incredibly useful for studies of stellar evolution and nucleosynthesis. This is because the stars are at a known distance from us, so we can estimate their absolute brightness and luminosities. Furthermore, because the AGB population is so large, they span the entire range of AGB initial masses, which evolve from stars of 1 to about 8$\Msun$, allowing us to place important constraints on the physics of this most uncertain phase of stellar evolution \citep[e.g.,][]{smith89,smith90b,plez93,marigo99,vanloon99a,vanloon99b,maceroni02,groen09,garcia09,gull12,kraemer17}. For reviews of the AGB phase of evolution we refer to \citet{karakas14dawes} and \citet{herwig05}. 

Many of the AGB stars in the Magellanic Clouds are in star clusters, allowing us to place firm constraints on their initial masses and ages \citep{kamath10,lebzelter07}, as well as compositions and evolutionary histories \citep{lebzelter08,lederer09b,kamath12}. Recent surveys of the Magellanic Clouds have revealed populations of post-AGB stars \citep{kamath14,kamath15}, which provide higher quality stellar abundance estimates than AGB stars because of their warmer atmospheres \citep{vanwinckel03,reyniers07a,desmedt12}. Finally, we can also use the large population of planetary nebulae (PNe) in the LMC and SMC for studies of stellar evolution and nucleosynthesis \citep[e.g.,][]{vw96,dopita97,stanghellini00,marigo03,leisy06,idiart07,bernard08,shaw10,ventura16b}. 

The present day interstellar medium of the Large and Small Magellanic Clouds have metallicities of approximately a factor of 2 and 5 below solar  \citep[e.g.,][]{russell92,gordon11}. This means that the metallicity of young stars are [Fe/H]\footnote{we use the standard spectroscopic notation, [A/B] = $\log_{10}(A/B)_{\rm surf} - \log_{10}(A/B)_{\odot}$. The ratio $(A/B)_{\rm surf}$ is the number ratio of elements $A$ and $B$ at the surface of the model star and $(A/B)_{\odot}$ is the solar number ratio, taken from \citet{asplund09}.} $=-0.3$ and [Fe/H] $=-0.7$ for the LMC and SMC. There are however large metallicity spreads in the stellar populations of the Magellanic Clouds. The median metallicity of the SMC is [Fe/H] $\approx -1$, with a low-metallicity tail extending to [Fe/H] $\approx -2$ \citep[e.g.,][]{parisi16}.

The aim of this paper is to present new stellar evolutionary sequences calculated with the Monash stellar evolution code, abundances and stellar yields from models of $Z=0.0028$ or [Fe/H] $\approx -0.7$ when adopting $Z_{\odot} = 0.014$ for the solar metallicity. This is a follow-up to \citet{karakas16}, where we present new stellar abundances and yields of AGB stars of solar metallicity and a factor of two above and below solar (e.g., [Fe/H] = $+0.3, 0, -0.3$). The new AGB models presented in this paper of [Fe/H] $=-0.7$ along with models presented in previous papers \citep{karakas16,fishlock14b,shingles15,lugaro12} span most of the range of metallicities of stars in the thin and thick disk of the Milky Way, and in the Large and Small Magellanic Clouds. 

The models presented here of [Fe/H] $= -0.7$ fill in the metallicity gap needed for studies of Galactic chemical evolution and the slow neutron capture process (the $s$ process). \cite{busso01} showed that the peak in [Ba/Fe] observed in AGB stars and related objects (e.g., barium stars, CH stars) lies at around [Fe/H] $= -0.7$ in the Galaxy. Previous models calculated at a similar metallicity (which adopted $Z=0.004$, based on a solar metallicity of $Z=0.02$) with the Monash stellar evolution code did not include a full $s$-process network, and presented surface abundances and yields of light-elements only \citep{karakas07b,karakas10a}. Here we fill in this gap by providing surface abundance predictions and yields for elements from C through to Bi.  

The outline of this paper is as follows. In Section~\ref{sec:agb} we provide a brief introduction to AGB stars and we introduce our methodology in Section~\ref{sec:method}. Section~\ref{sec:models} presents the new stellar evolutionary sequences and in Section~\ref{sec:nucleo} we present the stellar abundances and yields. In Section~\ref{sec:discuss} we compare our models to observations in the literature of a similar metallicity and discuss implications. We conclude in Section~\ref{sec:conclude}.

\section{AGB stellar models} \label{sec:agb}

Stars with initial masses between about 0.8 to 8$\Msun$, depending on metallicity, will evolve through core hydrogen and helium burning before ascending the AGB \citep{busso99,herwig05,karakas14dawes}. It is during the AGB phase that the richest nucleosynthesis occurs, driven by He-shell instabilities. These instabilities or thermal pulses (TP) may drive third dredge-up mixing, which occurs when material from the H-exhausted core is mixed into the envelope. TDU will alter the composition of the envelope by bringing the products of He-shell burning and neutron-capture nucleosynthesis to the stellar surface. Prior to the AGB, the first and second dredge-up may occur; these mix the products of hydrogen burning from the main sequence to the envelope.

Low-mass AGB stars with initial masses $M \lesssim 4\Msun$ have surface compositions that show enrichments in carbon, nitrogen, fluorine, and $s$-process elements \citep[e.g.,][]{busso01,karakas07a,weiss09,cristallo15,marigo17}. Intermediate-mass AGB stars on the other hand experience the second dredge-up (SDU) during the early AGB and hot bottom burning (HBB), the process by which the base of the envelope becomes hot enough for proton-capture nucleosynthesis \citep{karakas03b,ventura13}. The surface chemistry of intermediate-mass stars shows the signature of proton-capture nucleosynthesis \citep[e.g., nitrogen enhancements,][]{mcsaveney07}. The efficiency of third dredge-up and the contribution of He-shell nucleosynthesis to the surface chemistry of intermediate-mass stars is still debated \citep[e.g.,][]{karakas12,kalirai14,ventura15}. 

Here we evolve models with masses between 1$\Msun$ and 7$\Msun$ from the main sequence to near the tip of the AGB. The mass range includes the full range of CO-core AGB stars and one super-AGB star with an O-Ne core: the 7$\Msun$ model, which experiences off-centre carbon ignition  \citep[e.g.,][]{doherty14b}. 

\section{Methodology} \label{sec:method}

We employ the same methodology described in detail by \citet{karakas14b} and \citet{karakas16}. We first calculate stellar evolutionary sequences from the main sequence to the tip of the AGB using the Monash stellar evolution code \citep[][and references therein]{karakas14b}. These stellar evolution models do not include rotation or non-standard mixing phenomena beyond convective overshoot at the base of the convective envelope during the thermally-pulsing AGB (see paragraph below). Rotation in stellar codes has been shown to increase the growth of the H-exhausted core and can promote mixing of internal layers with the surface regions, as shown by e.g., \citet{decressin09}. Furthermore, rotation has been observationally discovered to be important in young LMC and SMC star clusters \citep{milone16}.

We calculate the low and intermediate-mass models with a scaled-solar composition and $Z=0.0028$, where we adopt $Z_{\odot}=0.014$ for the solar metallicity. This results in [Fe/H] = $-0.7$. Stars in the Milky Way Galaxy with a similar metallicity have some $\alpha$-enhancement \citep[e.g., see][]{reddy06} but usually only at a level of 0.2~dex. Furthermore, stars in the SMC with a similar metallicity of our models show either no or only a mild $\alpha$-enhancement \citep{russell92,mucc14}.  We employ exactly the same input physics in the stellar evolutionary sequences to \citet{karakas16} with the exception of AGB mass loss. 

Here we adopt the \citet{bloecker95} mass-loss law for intermediate-mass stars $\ge 4\Msun$ with $\eta=0.02$ \citep[which is the same as used by][]{ventura13}. The \citet{vw93} mass-loss law is used for lower mass models. We calculate evolutionary sequences for the 3.75, 4, 4.5, 5, 6 and 7$\Msun$ models with both the \citet{vw93} and \citet{bloecker95} mass-loss prescriptions in order to compare differences in the evolution and nucleosynthesis.

Our justification and reasons for these choices are as follows. \citet{groen09} present mass-loss rates and luminosities of AGB stars in the LMC and SMC, respectively. They find that the \citet{vw93} mass-loss law is a good approximation for the C-rich AGB stars but not for the intermediate-mass stars that experience HBB.  For this reason we decide to experiment with the AGB mass-loss law for models with HBB. Furthermore, we noted previously in \citet{karakas14b} when using the \citet{vw93} mass-loss law that lower metallicity models of $Z=0.007$ with HBB experience many more thermal pulses than models of lower mass, with no HBB. We can see no physical reason why this should be the case and suspect it is caused by our implementation of the low-temperature opacities (or the opacities themselves) used in the calculations. Other studies have highlighted the dependence of HBB on the low-temperature opacities \citep{ventura10,fishlock14a,constantino14} although further investigation is required to find out if our implementation of the low-temperature molecular opacities from \citet{marigo09} is indeed the cause. However the main physical reason is that our intermediate-mass models of lower metallicity stay compact for longer, which delays the onset of the superwind owing to the fact that the \citet{vw93} has a strong dependence on stellar radius.

We decide to adopt the \citet{bloecker95} mass-loss rate for intermediate-mass stars because it is widely used by other authors \citep[e.g.,][]{ventura13,pignatari16}. Our reasons are that \citet{ventura00} showed that the high fraction of Li-rich stars above $M_{\rm bol}=-6$ in the Magellanic Clouds demanded high mass-loss rates, and tentatively calibrated the free parameter $\eta$ in the \citet{bloecker95} formulation applied to their models. With our choice of $\eta = 0.02$ our intermediate-mass AGB models become rich in Li and remain so for $\approx 100,000$~years. The \citet{bloecker95} mass-loss rate is strongly dependent on the stellar luminosity, which will be lower in our AGB models than in the \citet{ventura13} models, which means that rates of mass-loss in the intermediate-mass AGB models will not be the same. Our AGB luminosities are lower because we adopt the Mixing-Length Theory of convection (with a mixing-length parameter, $\alpha = 1.86$) which has been shown to reduce the luminosity in AGB models with HBB \citep{mazzitelli99,ventura05a}.

The surface abundances and yields for the intermediate-mass models are calculated with the \citet{bloecker95} mass-loss except for the 3.75$\Msun$, 5$\Msun$ and 7$\Msun$ models, where we calculate nucleosynthesis for stellar evolutionary sequences calculated with  \citet{bloecker95} mass-loss and \citet{vw93} mass-loss to allow a comparison of the stellar yields.

Similar to \citet{karakas16} we include convective overshoot in low-mass AGB models (here $M=1.15, 1.25\Msun$) in order that they experience third dredge-up after thermal pulses on the AGB. We include overshoot because observations of carbon-rich AGB stars in the Magellanic Clouds suggest that the present-day AGB mass of the lowest mass C-stars is around 1$\Msun$ \citep[see Chapter 2 in][]{agbstars}. These observations can only be reproduced with the inclusion of overshoot with the Monash code. The treatment of overshoot in the $Z=0.0028$ models is exactly the same as described in \citet{karakas16} where we extend the base of the envelope by $N$ pressure-scale heights, where $N=1$ for both the 1.15$\Msun$ and 1.25$\Msun$ models.

For the detailed nucleosynthesis of elements between hydrogen through to bismuth, we use a post-processing code in the same manner to \citet{karakas16}. We employ the same nuclear network and initial solar abundances and we refer to that paper for full details. 

\subsection{The inclusion of $^{13}$C pockets}

AGB stars and their progeny are observed to be enriched in $s$-process elements by up to 1 dex at solar metallicity \citep[see e.g.,][]{busso01,abia02}. This implies that a large number of neutrons need to be released in the He-intershell by ($\alpha$,n) reactions. The main source of neutrons 
is the \iso{13}C($\alpha$,n)\iso{16}O reaction, which requires a reservoir of \iso{13}C in order to be activated. CNO cycling does not leave enough \iso{13}C nuclei in the He-intershell to produce $s$-process elements \citep[e.g.,][]{karakas07a}.  
The usual solution to this problem is to assume that some partial mixing of protons occurs between the convective envelope and the He-intershell. This mixing is assumed to occur at the deepest extent of each TDU episode where a sharp discontinuity is present between the H-rich convective envelope and the He-rich radiative intershell. The protons mixed into the top of the He-intershell are captured by \iso{12}C to produce a region rich in \iso{13}C, the so-called \iso{13}C ``pocket''. 

The formation of \iso{13}C pockets in theoretical calculations of AGB stars is one of the most significant uncertainties affecting predictions of the $s$ process \citep[see discussion in][]{busso99,herwig05,karakas14dawes}. A range of possibilities and models has been proposed to mix protons into the He-intershell at the deepest extent of each TDU episode. Among the most recent proposed mechanisms, \citet{herwig00} and \citet{cristallo09} employed convective overshoot \citep[see also][for a recent analysis]{goriely18}, \citet{denissenkov03a} internal gravity waves, and \citet{trippella16} magnetic fields. Here we adopt the same techniques we have applied before \citep{fishlock14b,karakas16} and outlined in detail in \citet{karakas16} and \citet{buntain17}. 

Briefly, we include \iso{13}C pockets by means of an artificial mixing profile driving the mixing of protons into the intershell. The mixing is modelled using an exponential function where the exponent is a linear function of the mass. This method produces $s$-process results for low-mass AGB stars very close to those produced by the overshoot models of \citet{cristallo09}, as discussed in detail in previous studies \citep{lugaro12,kamath12,fishlock14b,karakas16}. Here, we test different mass extents $M_{\rm mix}$ of the region affected by the mixing. Possible variations in the exponential function were analysed in detail by \citet{buntain17}, who concluded that, typically, changing the shape of the mixing function or the mass extent $M_{\rm mix}$ produce similar results: they change the absolute yield values but do not affect the relative distribution of the $s$-process elements.

As noted above, we do not include rotation in our models. Rotational mixing could strongly affect the $s$-process by mixing \iso{14}N, a neutron poison via the \iso{14}N(n,p)\iso{14}C reaction \citep{wallner16}, into the \iso{13}C pocket \citep{herwig03,siess04,piersanti13}. Investigations are currently under way to establish the strength of this rotational effect and the link with the astereoseismology observational constraints that the cores of giant stars and white dwarfs have lower rotation periods than expected by stellar models \citep{cantiello14}.

In Table~\ref{tab:c13} we show the entire range of stellar nucleosynthesis models calculated for each stellar mass, where the value of $M_{\rm mix}$ is chosen as function of the stellar mass. Table~\ref{tab:c13} shows that we include \iso{13}C pockets in all AGB models $M \le 4.5\Msun$ which experience TDU. Owing to variations in the distribution of $s$-process elements in stars, we experiment with varying the parameter, $M_{\rm mix}$, in a sample of the models as demonstrated by Table~\ref{tab:c13}.

For intermediate-mass AGB models above 4.5$\Msun$ we do not include a \iso{13}C pocket, which follows observational evidence that the \iso{13}C neutron source is not present in intermediate-mass AGB stars \citep{garcia13}. Theoretical models also support this choice \citep{goriely04} because the hot base of the envelope destroys protons during dredge-up, before they can be captured by \iso{12}C in the intershell.  The $s$-process predictions of \citet{cristallo15} show activation of the \iso{13}C neutron source in their intermediate-mass models, although, the signature at the stellar surface is very weak due to the low TDU efficiency. This could be because HBB is not very strong and protons are subsequently not destroyed during TDU. 

\begin{table*}
 \begin{center}
  \caption{The stellar nucleosynthesis models calculated for $Z=0.0028$: A tick (\checkmark) shows the size of $M_{\rm mix}$ used in the calculations. The [ST] label indicates the standard choice for each model. We also list the AGB mass-loss prescription used, where "VW93" refers to \citet{vw93} and "B95" \citet{bloecker95}.}
 \label{tab:c13}
  \vspace{1mm}
   \begin{tabular}{ccccccc}
   \hline\hline
 $M_{\rm mix}/\Msun=$ & Mdot & 0  & $1\times 10^{-4}$ & $1\times 10^{-3}$ &
 $2 \times 10^{-3}$ & $6 \times 10^{-3}$  \\ 
\hline
 Stellar Mass ($\Msun$) &   &  &   &   &  & \\
\hline \hline
 1.00 & VW93 & \checkmark &  &  &    &     \\
 1.15 & VW93 &           &  & \checkmark & \checkmark\,[ST] & \checkmark \\
 1.25 & VW93 &           &  &  & \checkmark\,[ST] & \checkmark \\
 1.50 & VW93 &           &  & \checkmark  & \checkmark\,[ST] & \checkmark \\
 1.75 & VW93 &           &  &  & \checkmark\,[ST] &  \\
 2.00 & VW93 &           &  & \checkmark  & \checkmark\,[ST] & \checkmark \\
 2.25 & VW93 &           &  &  & \checkmark\,[ST] &  \\
 2.50 & VW93 &           &  &  & \checkmark\,[ST] & \checkmark$^{\rm a}$\\
 2.75 & VW93 &           &  &  & \checkmark\,[ST] &  \\ 
 3.00 & VW93 &           &  & \checkmark\,[ST] & \checkmark &   \\
 3.25 & VW93 &           &  & \checkmark\,[ST] &  &   \\
 3.50 & VW93 &           &  & \checkmark\,[ST] &  &  \\
 3.75 & VW93 &           & \checkmark  & \checkmark\,[ST] &  &  \\
 3.75 & B95  &           &             & \checkmark &  & \\
 4.00 & B95  &           & \checkmark\,[ST] & \checkmark  &  &  \\
 4.50 & B95  & \checkmark\,[ST] & \checkmark &  &  &  \\
 5.00 & B95  & \checkmark\,[ST] &  & &  &   \\
 5.00 & VW93 & \checkmark\,[ST] &  & &  &   \\
 5.50 & B95  & \checkmark\,[ST] &  & &  &     \\
 6.00 & B95  & \checkmark\,[ST] &  &  & &     \\
 6.50 & B95  & \checkmark\,[ST] &  &  & &     \\
 7.00 & B95  & \checkmark\,[ST] &  &  & &     \\
 7.00 & VW93 & \checkmark\,[ST] &  &  & &     \\
\hline \hline
  \end{tabular} 
\\
(a) -- in this case we used $M_{\rm mix} = 4\times 10^{-3}\Msun$\\
\end{center}
\end{table*}

\section{AGB model results}  \label{sec:models}

In this section we focus on structural details of the AGB models that are relevant for the nucleosynthesis and therefore shape the stellar yields. These include the H-exhausted core mass at the beginning of the thermally-pulsing AGB, the number of thermal pulses on the AGB, the amount of material dredged up to the surface, along with the temperature of the hydrogen and helium shells and the temperature at the base of the convective envelope. 

\begin{figure}
    \centering 
    \includegraphics[width=0.95\columnwidth]{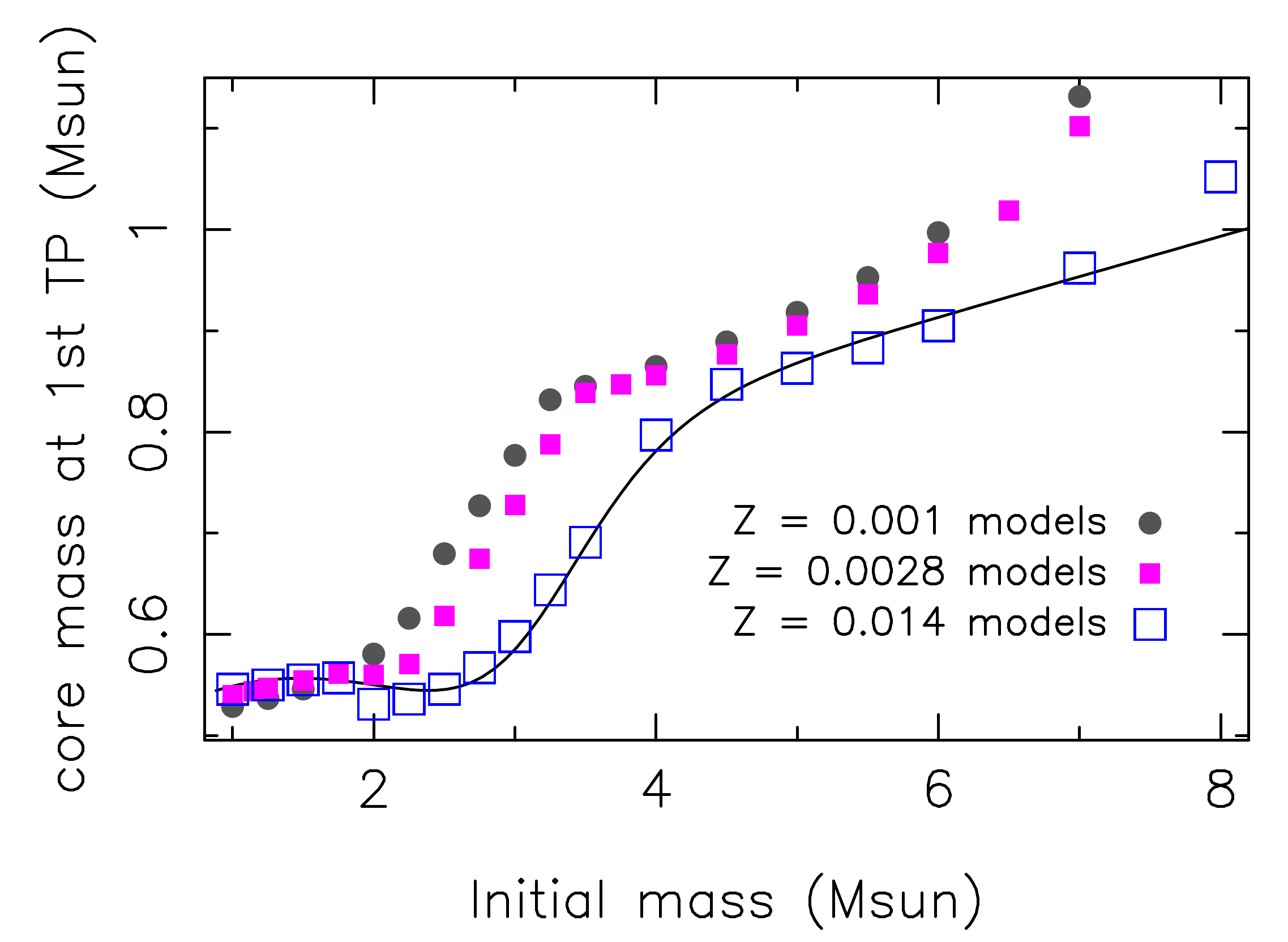}
    \caption{The core mass at the beginning of the thermally-pulsing AGB for the $Z=0.0028$ models, which we take at the start of the first thermal pulse. We also include the $Z=0.001$ models from \citet{fishlock14b} and the solar-metallicity $Z=0.014$ models from \citet{karakas14b}. The solid line shows the parameterized fit to the core mass at the first thermal pulse from \citet{karakas02} for models of $Z=0.02$.}
    \label{fig:mc1}
\end{figure}

In Fig.~\ref{fig:mc1} we show the H-exhausted core mass (hereafter core mass) at the beginning of the thermally-pulsing AGB for the $Z=0.0028$ models. We compare the core mass at this stage in the evolution to models of $Z=0.001$ from \citet{fishlock14b}, which while slightly lower metallicity are very similar to the models presented here. For comparison we show the core mass at the first thermal pulse for models of solar metallicity ($Z=0.014$), noting that the ``knee" around 4$\Msun$ in the solar-metallicity models shifts to $\approx 3\Msun$ at lower metallicities. This means that the minimum mass for HBB, which strongly depends on this parameter, will shift to a lower initial mass. Note that we find no difference in core mass at the beginning of the AGB for models calculated with \citet{vw93} mass-loss compared to models calculated with \citet{bloecker95} mass-loss.

\begin{figure}
    \centering 
    \includegraphics[width=0.95\columnwidth]{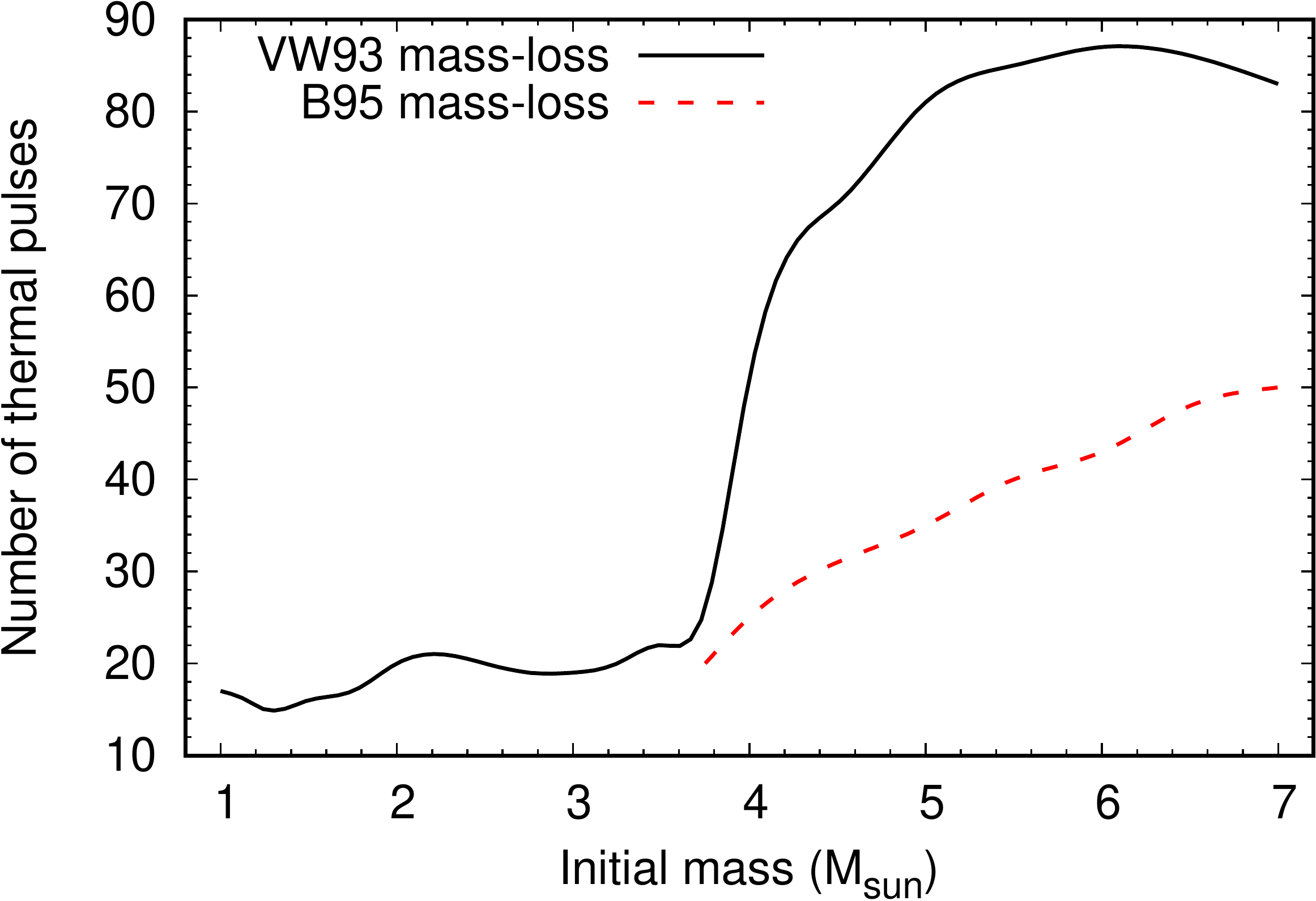}
    \caption{Number of thermal pulses calculated for the AGB models. The black solid line shows AGB models calculated with \citet{vw93} mass-loss while the red dashed line shows AGB models calculated with  \citet{bloecker95} with $\eta = 0.02$.}
    \label{fig:tps}
\end{figure}

In Fig.~\ref{fig:tps} we show the number of thermal pulses from each calculation as a function of the initial mass.  Results from models calculated with \citet{vw93} mass-loss on the AGB are shown by the solid line while results for intermediate-mass models calculated with \citet{bloecker95} are shown by the dashed line. \cite{vw93} mass-loss results in $\lesssim 30$~TPs for models with $M < 4\Msun$ but that number greatly increases in intermediate-mass models which experience efficient hot bottom burning. Fig.~\ref{fig:tps} shows that models with \citet{vw93} experience roughly twice as many thermal pulses on the AGB compared to models calculated with \citet{bloecker95} mass-loss with $\eta = 0.02$. 

\begin{figure}
    \centering 
    \includegraphics[width=0.95\columnwidth]{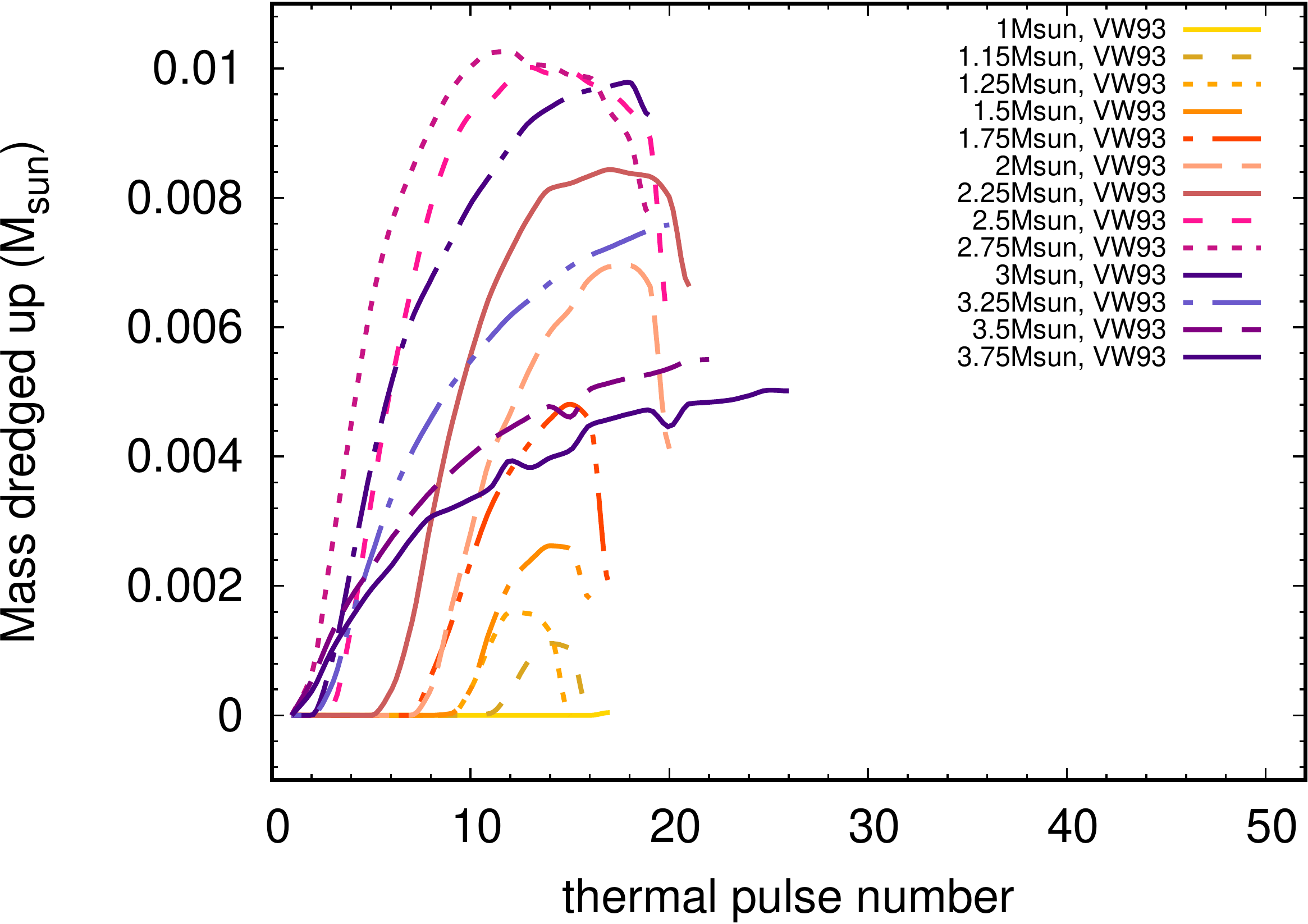}
    \includegraphics[width=0.95\columnwidth]{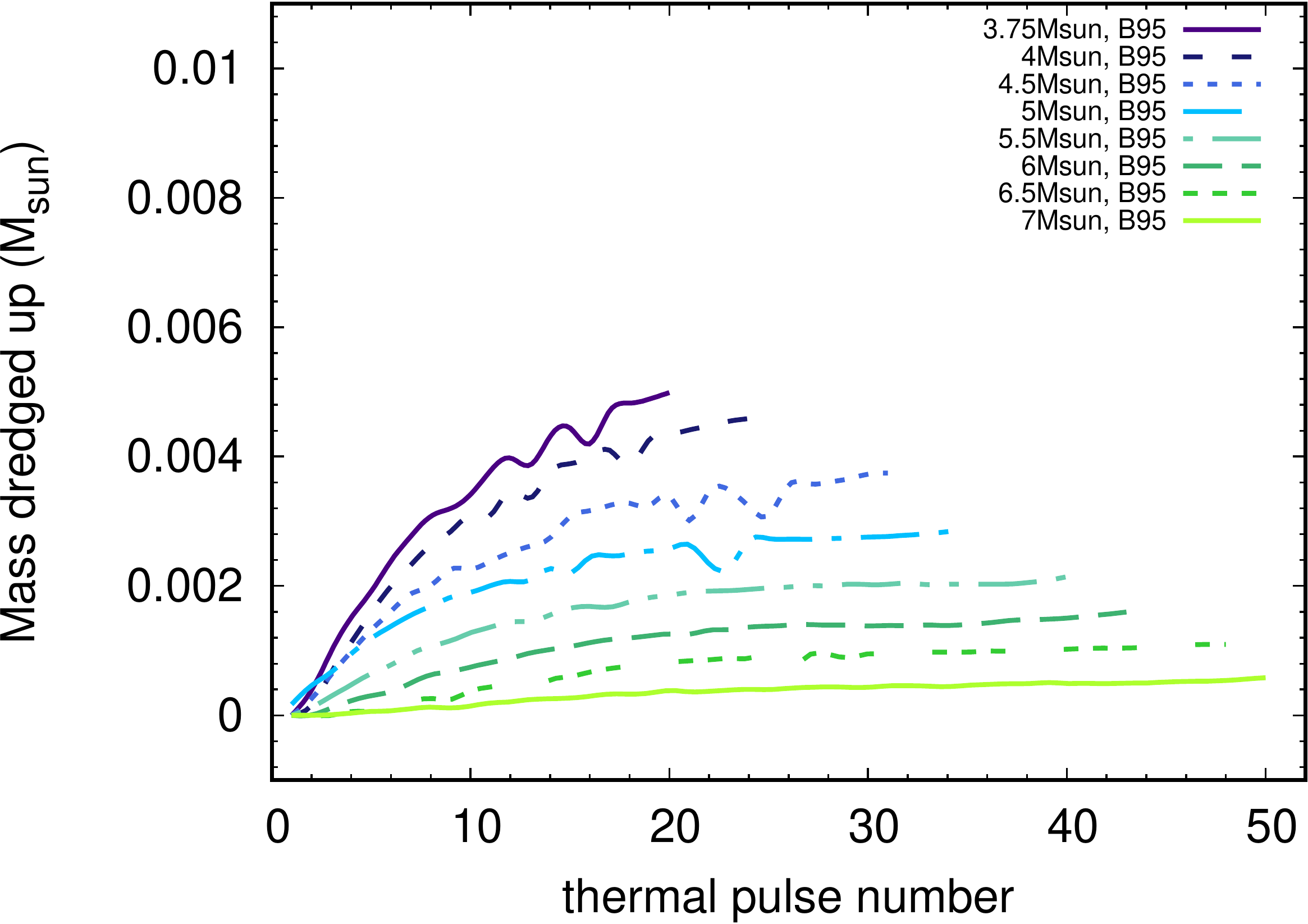}
    \caption{Mass dredged-up during each thermal pulse for the $Z=0.0028$ models. The upper panel shows low-mass models calculated with \citet{vw93} mass-loss on the AGB (VW93 in the legend); the lower panel intermediate-mass models calculated with \citet{bloecker95} (B95 in the legend). We include the 3.75$\Msun$ model results in both panels for comparison.} 
    \label{fig:mdredge}
\end{figure}

The amount of material dredged-up from the He-intershell is important for determining the surface composition during the AGB and the final yields. In Fig.~\ref{fig:mdredge} we show the mass dredged-up as a function of thermal pulse for each of the $Z=0.0028$ models for which we also present stellar yields. We do not show the amount of mass dredged up for the intermediate-mass models calculated with \citet{vw93} mass-loss but as Fig.~\ref{fig:tps} illustrates, these models experience roughly twice as many thermal pulses and consequently dredge-up roughly twice as much.  Using the 6$\Msun$ model as an example, the model with \citet{vw93} dredges up 0.105$\Msun$ in total from the He-intershell, roughly twice that of the model calculated with \citet{bloecker95} mass-loss which dredges up 0.0469$\Msun$. 

\begin{figure}
    \centering 
    \includegraphics[width=0.95\columnwidth]{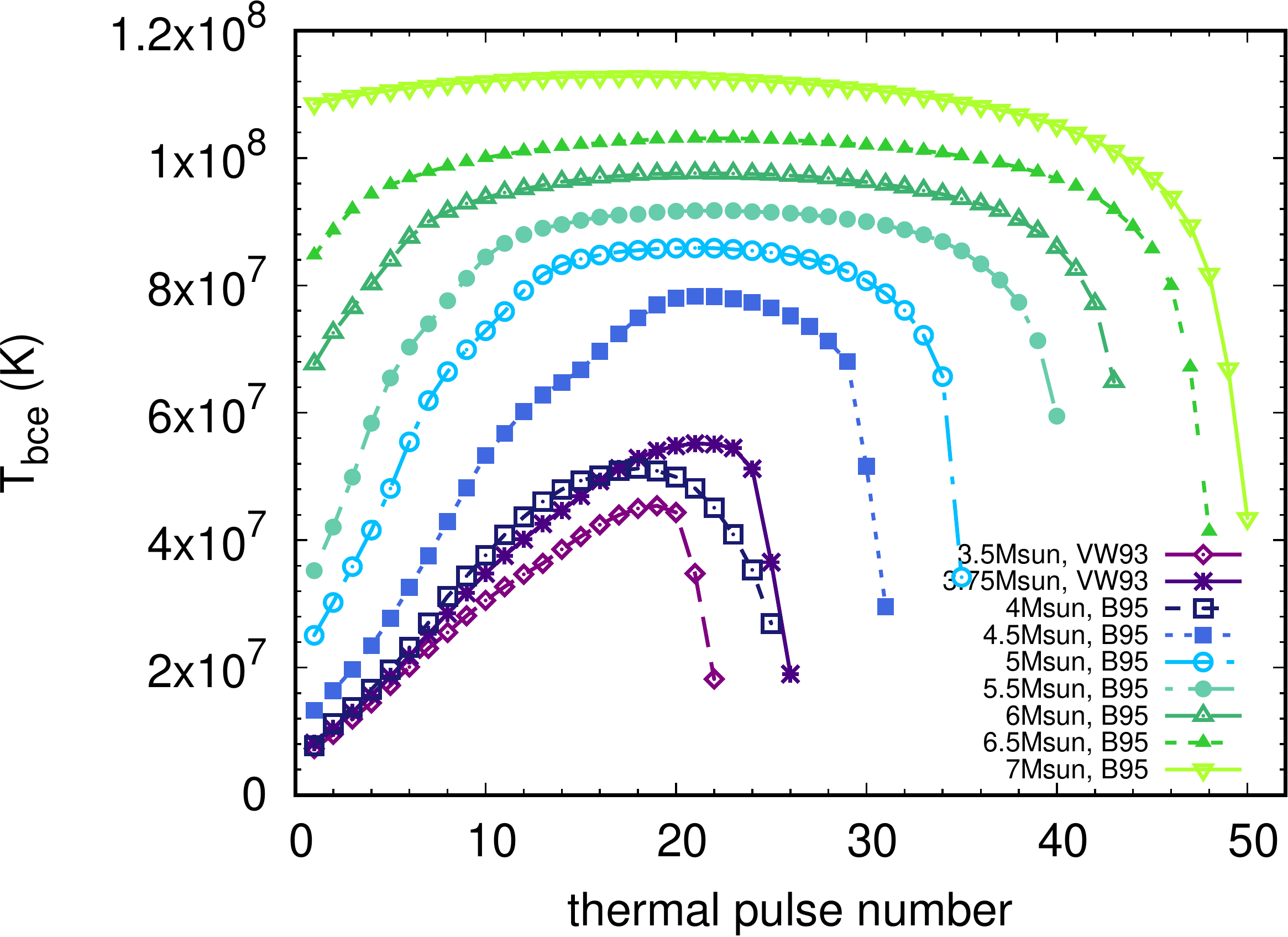}
    \caption{Maximum temperature at the base of the envelope during the interpulse phase as a function of thermal pulse number for models $\ge 3.5\Msun$.}
    \label{fig:hbb}
\end{figure}

Fig.~\ref{fig:hbb} shows the maximum temperature at the base of the envelope predicted for the intermediate-mass models ($M \ge 3.5\Msun$). When the temperature at the base of the envelope exceeds  about $50\times 10^{6}$~K (hereafter MK; where the exact minimum mass depends also on the metallicity and hence structure of the envelope), proton-capture nucleosynthesis or HBB can occur. For the $Z=0.0028$ the minimum stellar mass for HBB with the Monash stellar evolution code is 3.75$\Msun$, which has a maximum HBB temperature of 55~MK. 

The minimum mass depends on the mass-loss used on the AGB. With \citet{bloecker95} mass-loss the minimum initial mass increases to 4$\Msun$ from 3.75$\Msun$. This is because models with \citet{bloecker95} mass-loss lose their envelope masses faster, which consequently reduces the temperature at the base of the envelope. Fig.~\ref{fig:hbb} shows that the 3.75$\Msun$ model with \citet{vw93} mass-loss reaches a higher maximum temperature than the 4$\Msun$ model with \citet{bloecker95} mass-loss on the AGB.  To show just how much the maximum temperature depends on mass-loss, the 4$\Msun$ model with \citet{vw93} has a maximum HBB temperature of 78.7~MK, about 50\% higher, as a consequence of the envelope mass remaining higher for longer.

The minimum initial mass for HBB is also strongly dependent upon the treatment of convection in stellar envelopes as highlighted by \citet{ventura05a} and subsequent papers \citep[e.g.,][]{ventura15}. The ATON code finds the minimum mass for HBB to be about 1$\Msun$ lower than models calculated with the Monash stellar evolution code. The FRUITY models by \citet{cristallo15} on the other hand find much milder HBB compared with the Monash models as discussed by those authors and \citet{karakas16}. The MESA/NuGrid models of \citet{pignatari16} predict HBB at a similar minimum mass to the Monash models, while the models by \citet{weiss09} finds that HBB occurs 1$\Msun$ higher (e.g., 6$\Msun$ for solar metallicities whereas we find efficient HBB at 5$\Msun$). \citet{colibri} obtains higher luminosities than models calculated with the Monash code, again highlighting the sensitivity of HBB to the input physics. 

\begin{figure}
    \centering 
    \includegraphics[width=0.95\columnwidth]{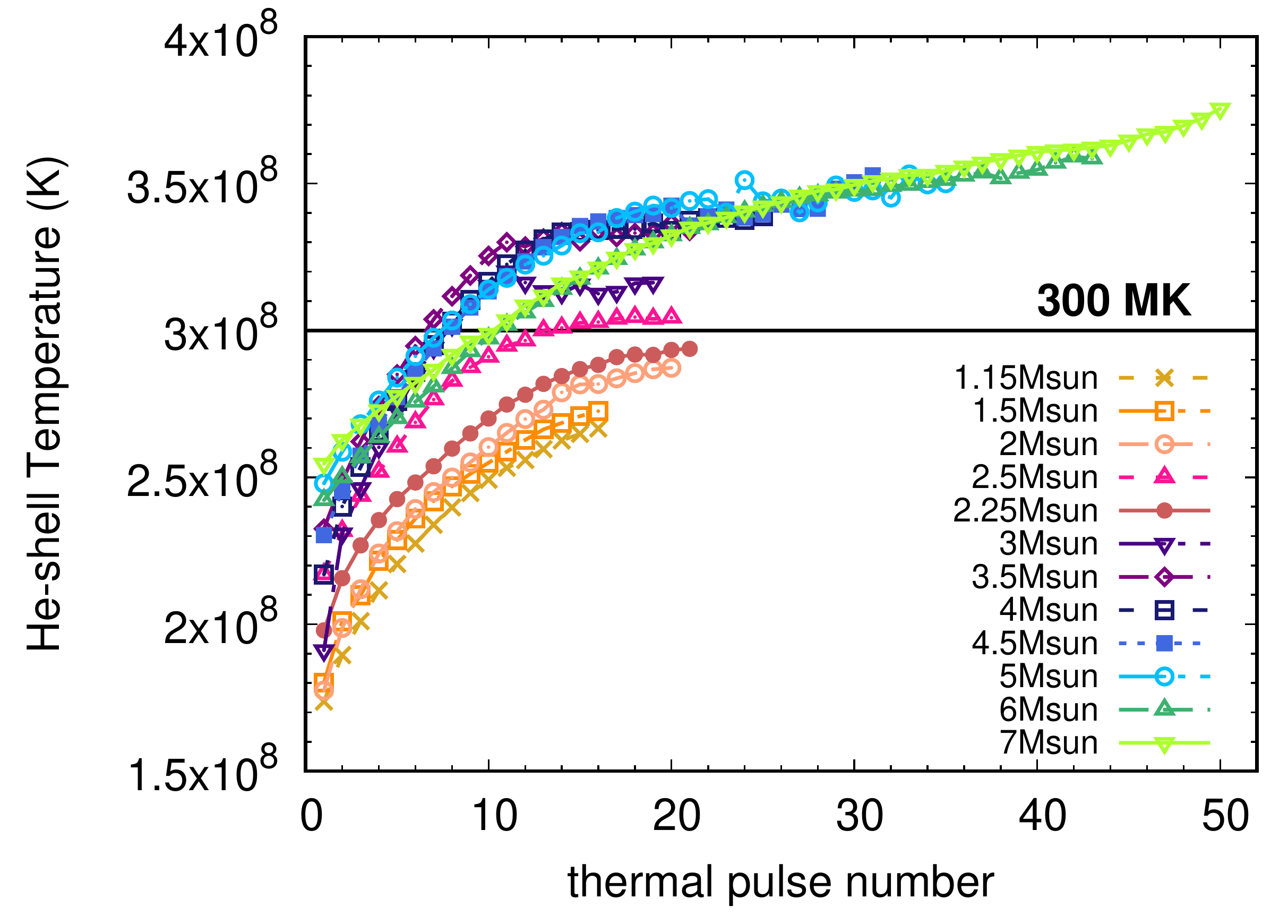}
    \caption{Maximum temperature during He-shell flashes as a function of thermal pulse number for a selection of the $Z=0.0028$ models.}
    \label{fig:heshell}
\end{figure}

Fig.~\ref{fig:heshell} shows the maximum temperature in the He-shell during thermal pulses for a selection of the $Z=0.0028$ models. We also include a line at 300~MK to show which models fall below and above, in terms of maximum He-shell temperatures. Models less than 2.5$\Msun$ do not reach peak He-shell temperatures above 300~MK, which means that the \iso{22}Ne($\alpha$, n)\iso{25}Mg neutron source is never activated. For these low-mass models, neutrons are only released by the \iso{13}C($\alpha$, n)\iso{16}O reaction. For models $\ge 2.5\Msun$, He-shell temperatures exceed 300~MK, which means that there will be a burst of neutrons released at high density during thermal pulses. Fig.~\ref{fig:heshell} illustrates that the 2.5$\Msun$ model only reaches temperatures of 300~MK during the last few TPs while for intermediate-mass models over 4$\Msun$ the majority of thermal pulses exceed 300~MK.  Consequences of the \iso{22}Ne $+ \alpha$ reactions include the production of \iso{25}Mg and \iso{26}Mg \citep[e.g.,][]{karakas03a} and the production of Rb over Sr, Zr \citep[e.g.,][]{vanraai12,karakas12}.

\section{Nucleosynthesis results} \label{sec:nucleo}

Here we present the results of the post-processing nucleosynthesis calculations.

\subsection{Data tables}
 
Similar to \citet{karakas16} we provide on-line only data tables for download. There are three tables: 1) surf\_z0028.dat, the elemental surface abundances as a function of thermal pulse number for all models shown in Table~\ref{tab:c13}; 2) isotope\_z0028.dat, the isotopic ratios of the elements up to Ni as a function of thermal pulse number; and 3) yields\_z0028.dat, the integrated elemental yields. In this section we summarize the nucleosynthesis results and stellar yields. In Appendix~\ref{app:datafiles} we provide examples of each of the datafile types available for download. 

The surface abundance data tables start with the initial abundances used in the post-processing calculations and then include elemental abundances as a function of thermal pulse number.  At each entry we include the thermal pulse number, the stellar mass, core mass and envelope mass at that thermal pulse (in $\Msun$), and the surface luminosity (in $\log \Lsun$). After the abundances of each element are given, with one row per element, we then provide the ratios of He/H, C/O and N/O at that thermal pulse.  The final entry for each ($M, Z$) combination are the final surface abundances, taken at the last time step, which may fall on a thermal pulse or during the interpulse period.

Our surface abundance data tables (both elemental and isotopic) are in exactly the same format as in \citet{karakas16}. This means that for all elements except Li, Be and B we include the element name, the proton number, $Z$; the abundance in the format $\log \epsilon (X)$ where $\log \epsilon (X) = \log_{10}(X/H) + 12$; [X/H], [X/Fe], [X/O], and the mass fraction $X(i)$. The radioactive elements Tc and Pm may be non-zero in terms of $\log \epsilon (X)$, if they are produced by neutron captures and mixed to the surface. 

We do not decay the abundances of  radioactive isotopes, because key isotopes such as \iso{26}Al and \iso{60}Fe, are produced in intermediate-mass stars and are observed in the isotopic grain data.  Radioactive isotopes are assumed to have all decayed in the yield tables. In the surface abundance tables we do decay \iso{93}Zr to \iso{93}Nb because the element Nb obtains most of its production by this decay.

In the isotopic data tables we include the following ratios:
\iso{12}C/\iso{13}C, \iso{14}N/\iso{15}N, \iso{16}O/\iso{17,18}O, 
\iso{24}Mg/\iso{25,26}Mg, \iso{26}Al/\iso{27}Al, \iso{28}Si/\iso{29,30}Si,
\iso{36,37}Cl/\iso{35}Cl, \iso{36,38}Ar/\iso{40}Ar, \iso{40,41}K/\iso{39}K, 
\iso{42,43,44,46,48}Ca/\iso{40}Ca, \iso{46,47,49,50}Ti/\iso{48}Ti,
\iso{53,54}Cr/\iso{52}Cr, \iso{54,57,58,60}Fe/\iso{56}Fe, and 
\iso{60,61,62}Ni/\iso{58}Ni.  

The integrated elemental yields are in a similar format to those available in \citet{karakas16} with the addition of the net yield, which is calculated according to
\begin{equation}
 M_{i} = \int_{0}^{\tau} \left[ X(t)_{\rm i} - X(0)_{\rm i} \right] \frac{d M}{dt} dt,  \label{eq:yield}
\end{equation}
where $M_{i}$ is the net yield of species $i$ (in solar masses), $dM/dt$ is the current mass-loss rate, $X(t)_{\rm i}$ the current mass fraction of species $i$ at the surface, $X(0)_{\rm i}$ is the initial mass fraction, and $\tau$ is the stellar lifetime. The net yield can be negative, when the element $i$ is destroyed by stellar evolution and mixing processes (e.g., hydrogen is only consumed). We also provide the total mass expelled which is calculated according to
\begin{equation}
 M_{i} = \int_{0}^{\tau} X(t)_{\rm i} \frac{d M}{dt} dt.  \label{eq:yield2}
\end{equation}
The yields from Equation~\ref{eq:yield2} are the total amount of element $i$ expelled into the interstellar medium over the stellar lifetime (in $\Msun$) and are always positive. 

In Table~\ref{tab:exampleyield} we show the first few lines of the yield tables for the 2$\Msun$, $Z = 0.0028$ model. We include the initial and final mass along with the total expelled mass (1.341$\Msun$ in this case). The header for each $(M,Z)$ combination also includes the $M_{\rm mix}$ used in the $s$-process calculations. The columns include the element name, proton number $Z$, the average abundance in the wind ejected in the following formats: 1) $\log \epsilon(X)$ which is defined according to $\log \epsilon(X) = \log_{10} (X/H) + 12$, where $X/H$ is the ratio of element $X$ to hydrogen (by number); 2) [X/H]; 3) [X/Fe]; 4) $X(i)$, which is the average mass fraction in the wind; 5) Net $M(i)$ calculated according to Equation~\ref{eq:yield}; and 6) the total mass of element $i$ expelled calculated according to Equation~\ref{eq:yield2}.

For nucleosynthesis models where we calculate AGB evolution with both the \citet{vw93} and \citet{bloecker95} mass-loss rates, we add to the header file either "VW93" or "B95" to 
indicate the mass-loss rate used.  

\subsection{Surface abundances on the AGB}

\begin{figure}
    \centering 
    \includegraphics[width=0.95\columnwidth]{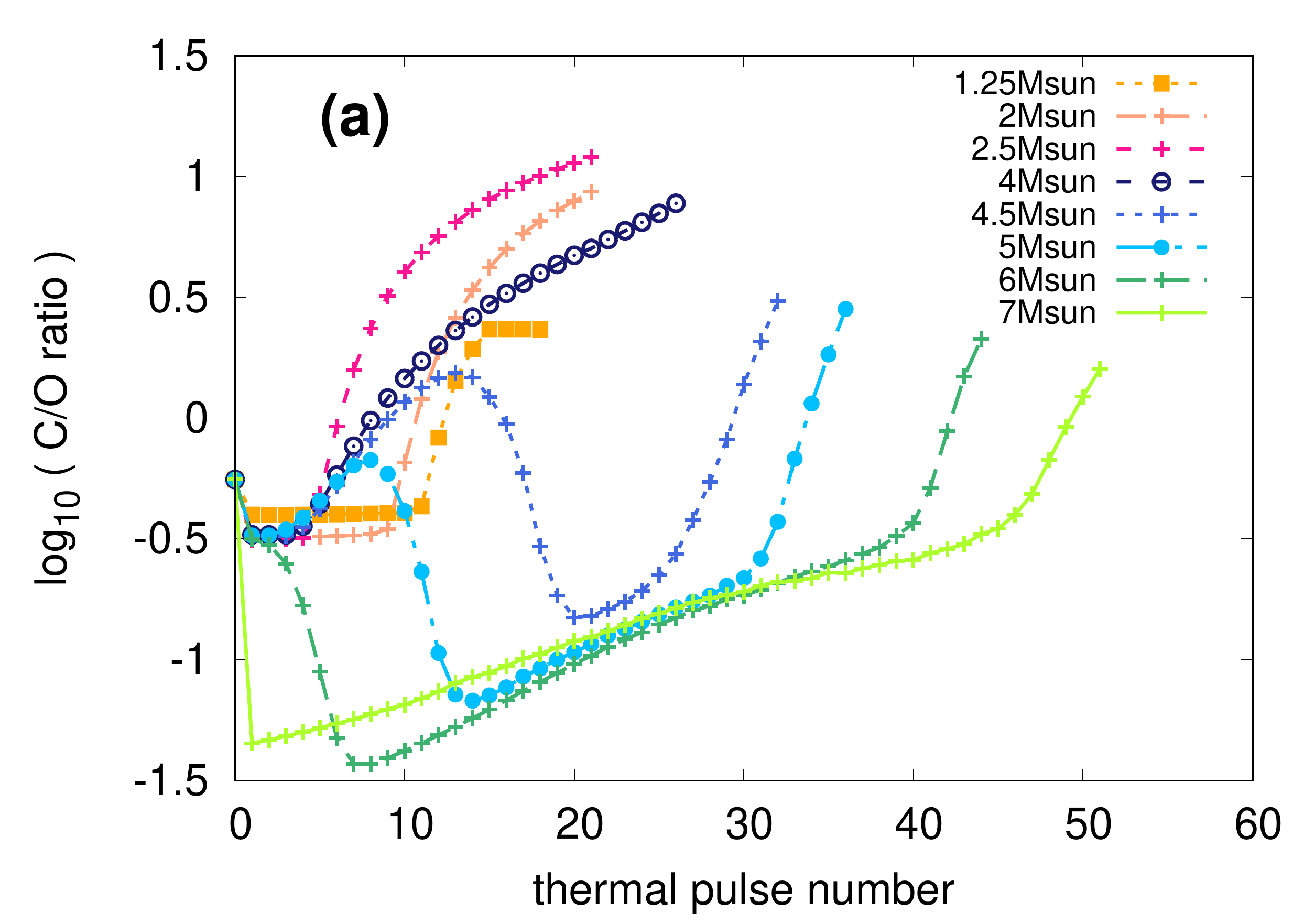}
    \includegraphics[width=0.95\columnwidth]{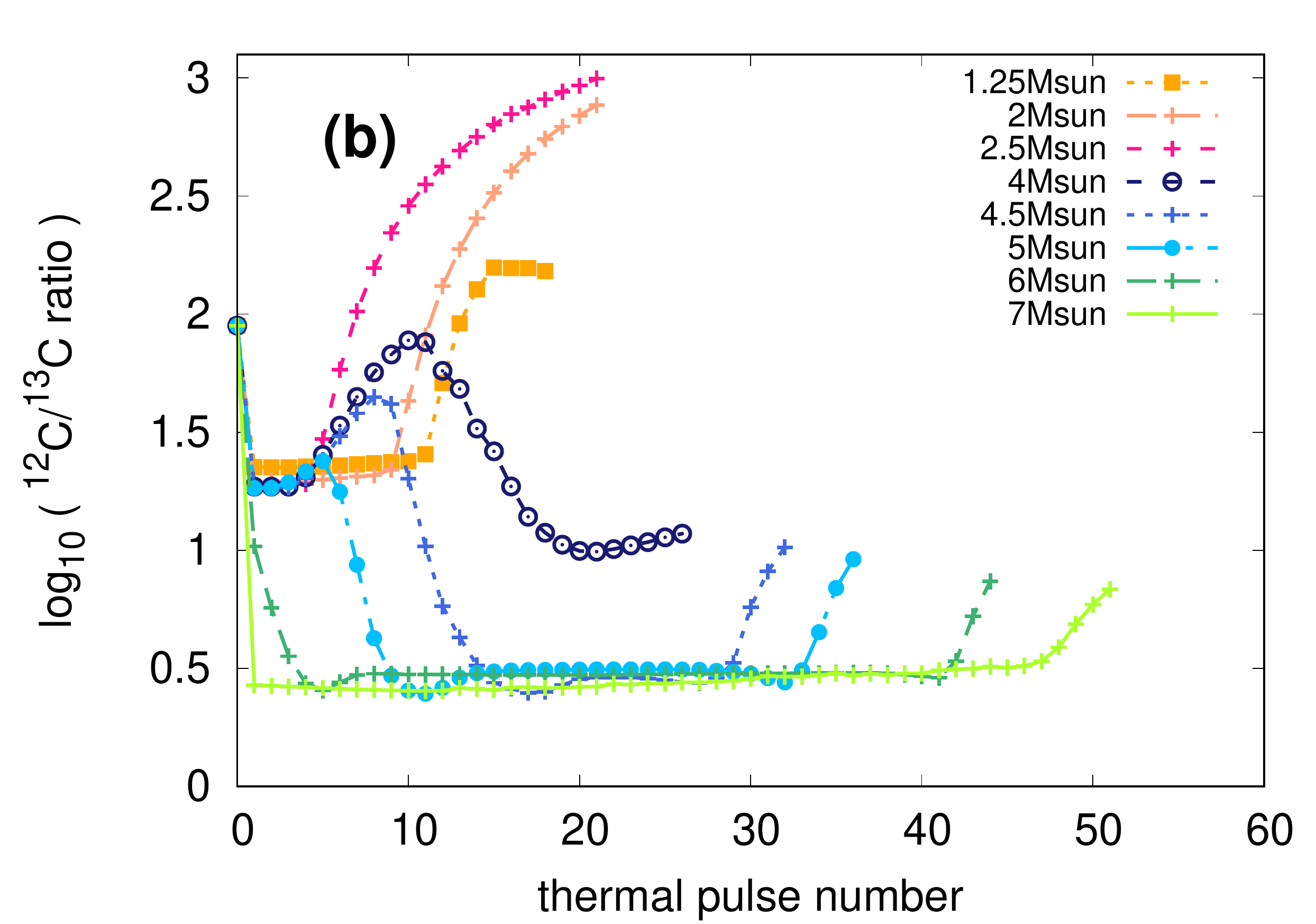}
    \caption{Log of the (a) surface C/O and (b)  \iso{12}C/\iso{13}C ratios as a function of thermal pulse number for a selection of $Z=0.0028$ models. All models start with the initial ratios which are solar: C/O = 0.55 ($\log_{10}$ C/O = $-0.26$) and \iso{12}C/\iso{13}C = 89 ($\log_{10}$ \iso{12}C/\iso{13}C = 1.95).}
    \label{fig:co}
\end{figure}

Here we summarize the evolution of the surface composition on the AGB. In the top panel of Fig~\ref{fig:co} we show the evolution of the C/O ratio as a function of thermal pulse number for a selection of the $Z=0.0028$ models, while the bottom panel of Fig.~\ref{fig:co} shows the evolution of the \iso{12}C/\iso{13}C ratio for the same set of models.

Fig.~\ref{fig:co} illustrates the complex behaviour of the C/O and \iso{12}C/\iso{13}C ratios in models with both TDU and HBB. For intermediate-mass models with efficient HBB, the C/O $<1$ during most of the thermally-pulsing AGB and it is only toward the end when sufficient mass-loss occurs that HBB is shut off and the C/O ratio increases and eventually exceeds unity. 

\begin{figure*}
    \centering 
\includegraphics[width=0.95\columnwidth]{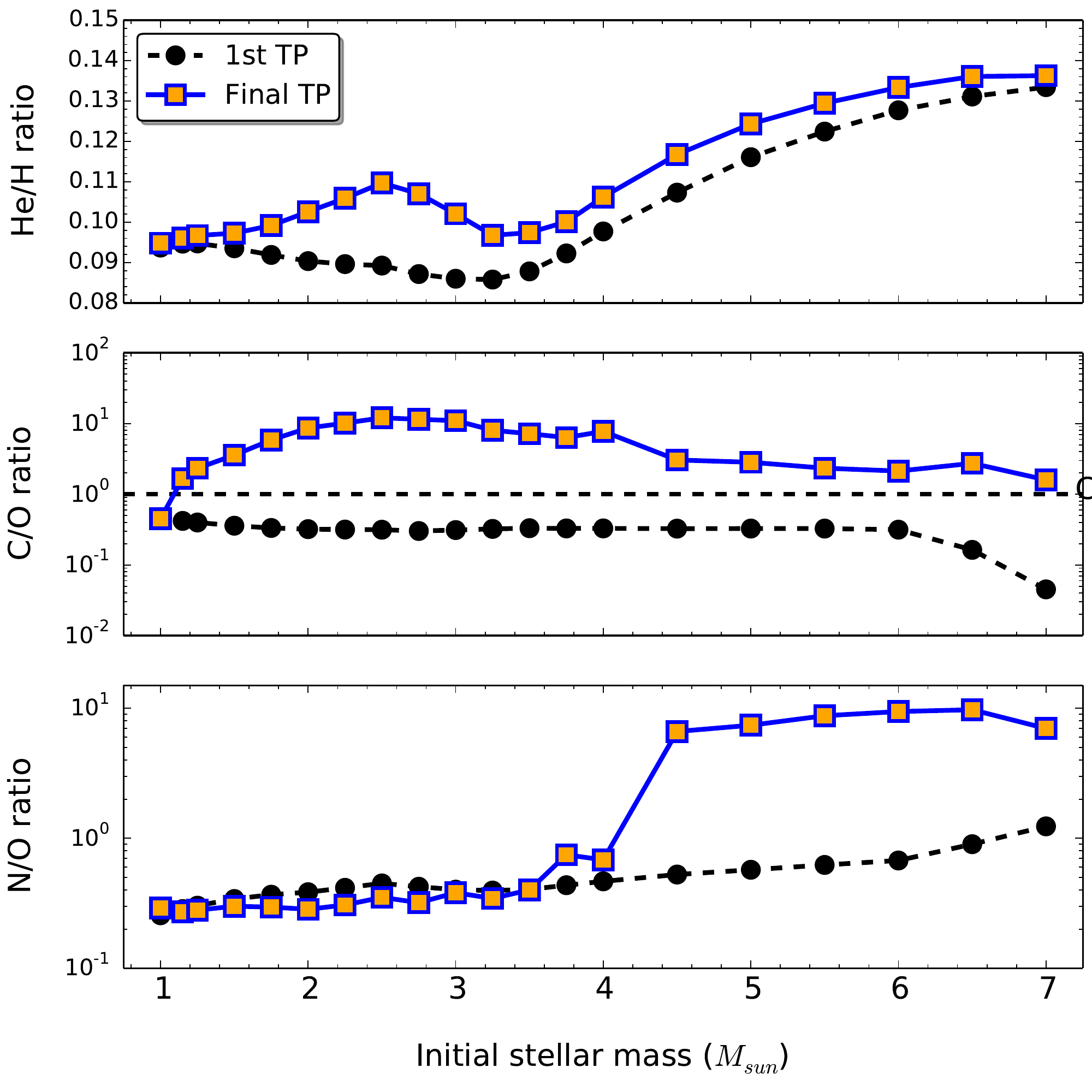}
  \caption{The He/H, C/O and N/O ratios at the beginning and end of the thermally-pulsing AGB phase for all models.}
   \label{fig:ratio1}
\end{figure*}

\begin{figure*}
    \centering 
\includegraphics[width=0.95\columnwidth]{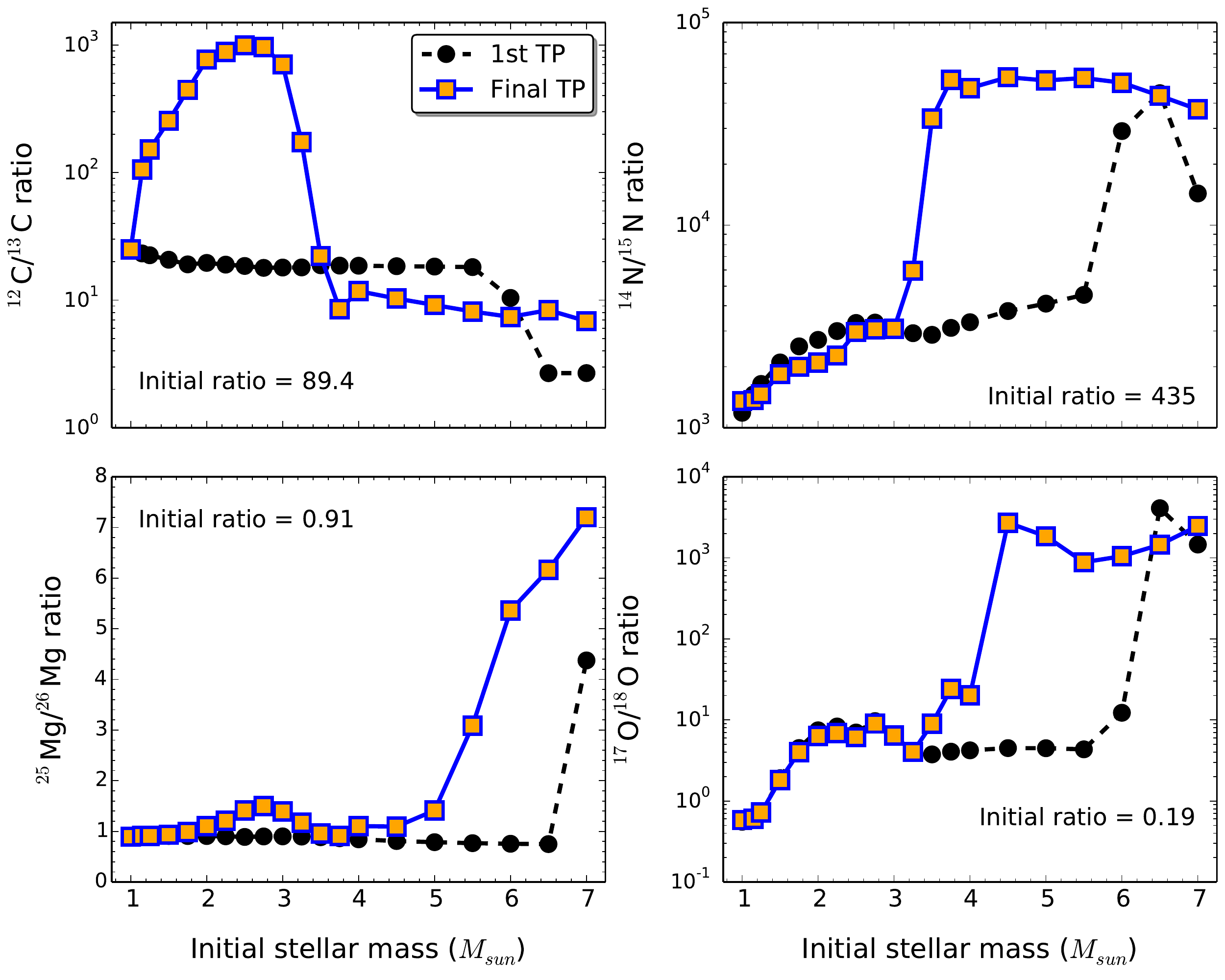}
  \caption{The \iso{12}C/\iso{13}C, \iso{14}N/\iso{15}N, \iso{25}Mg/\iso{26}Mg and \iso{17}O/\iso{18}O ratios at the beginning and end of the thermally-pulsing AGB phase for all models.}
   \label{fig:ratio2}
\end{figure*}

In Fig.~\ref{fig:ratio1} and~\ref{fig:ratio2} we show the ratios for He/H, C/O and N/O and \iso{12}C/\iso{13}C, \iso{14}N/\iso{15}N, \iso{25}Mg/\iso{26}Mg and \iso{17}O/\iso{18}O at the stellar surface. We include the ratios at the first thermal pulse and at the end of the AGB, at the last calculated time step. For most masses the ratios at the first thermal pulse reflect the post-first and second 
dredge-up abundances; the exceptions are the 6.5$\Msun$ and 7$\Msun$ models which begin HBB before the first thermal pulse. This is clear from the high N/O and \iso{14}N/\iso{15}N ratios which shows considerable CN processing, and from the low \iso{12}C/\iso{13}C ratio already evident at that stage. 

\begin{figure}
    \centering 
\includegraphics[width=0.95\columnwidth]{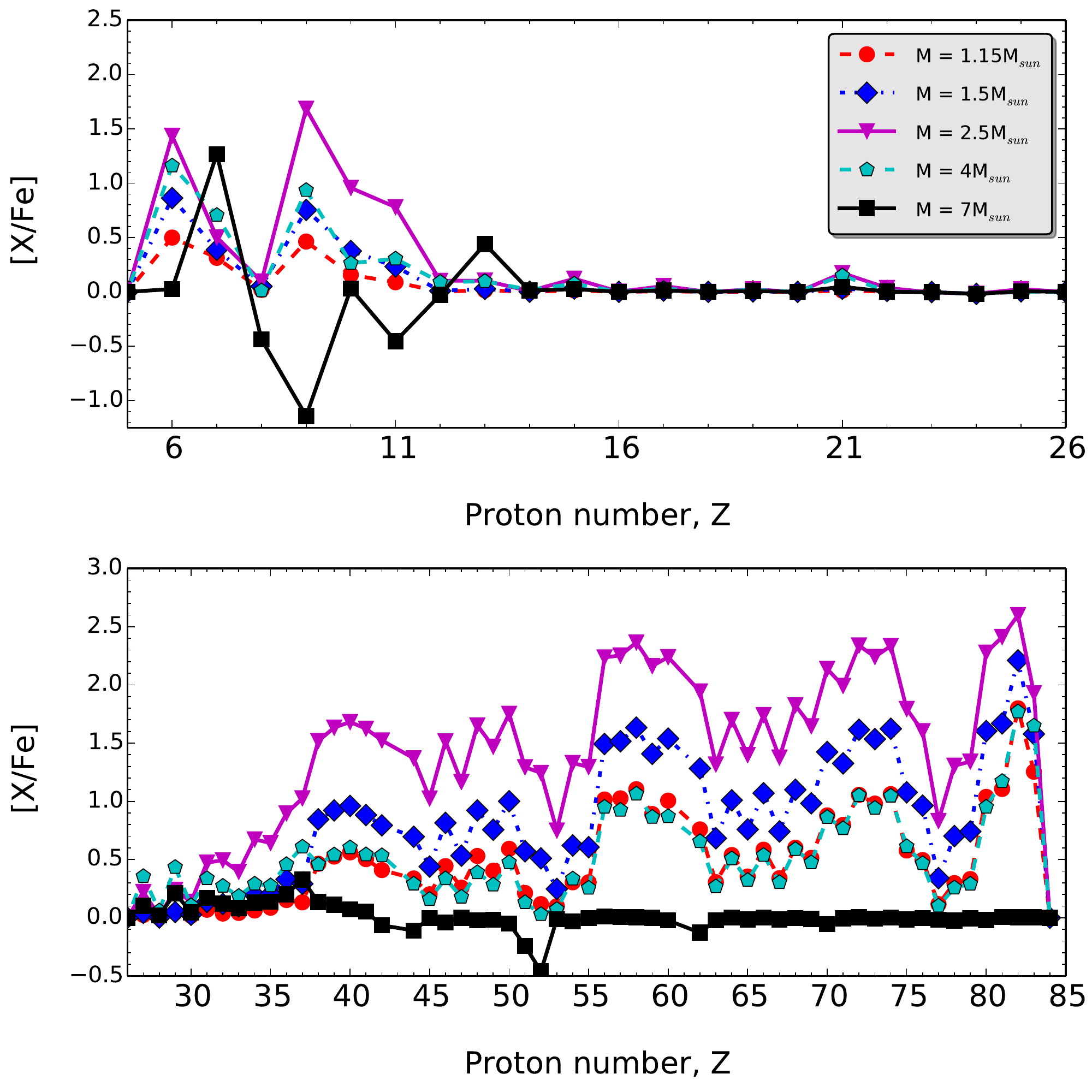}
  \caption{The final surface abundances (in [X/Fe]) plotted as a function of proton number, $Z$, for a selection of the $Z=0.0028$ models. We show in the upper panel the elements up to iron ($Z=26$) and in the lower panel for elements heavier than iron. We show the 4$\Msun$ model with a partial mixing zone, noting that the model with no \iso{13}C pocket has very little heavy-element production, e.g., as shown in Table~\ref{tab:sinds}. In other cases we choose the {\em standard} pocket size as indicated in Table~\ref{tab:c13}.}
   \label{fig:surf}
\end{figure}

In Fig.~\ref{fig:surf} we show the final surface abundances for a selection of the $Z=0.0028$ models between 1.15$\Msun$ and 7$\Msun$. The top panel shows elements lighter than iron and the lower panel elements heavier than iron. The 7$\Msun$ experiences strong HBB and we can see considerable destruction of some light elements as a result of CNO cycling (e.g., the final [O/Fe] $\approx -0.45$), along with strong N production. There is very little production of neutron-capture elements at the surface of the 7$\Msun$ model, even though temperatures in the He-shell peak at 376~MK, strong enough to activate the \iso{22}Ne($\alpha$,n)\iso{25}Mg reaction. An examination of the intershell during the final shell flash reveals strong enhancements of elements at the first $s$-process peak including Sr, Rb and Y. The model does not dredge-up much material from the He-intershell (0.0176$\Msun$ in total over the whole AGB) and the massive envelope results in a strong dilution. In comparison, the 3$\Msun$ model dredges up a total of 0.124$\Msun$ over the entire AGB. 

Fig.~\ref{fig:surf} shows that final surface abundance pattern for the lower mass models shows a strong  enrichment in elements typically produced by AGB stars including C, F and $s$-process elements. For the metallicity of [Fe/H] $=-0.7$ we see that the ratio of the second $s$-process peak around Ba, La, and Ce dominates, although a similar amount of Pb production also occurs.  In Table~\ref{tab:sinds} we show the $s$-process indicators defined according to \citet{lugaro12} for each calculation. We show [Rb/Zr], [ls/Fe] = (Sr+Y+Zr)/3, [hs/Fe] = (Ba+La+Ce)/3, [hs/ls], and [Pb/hs] and list the mass-loss formula used on the AGB and the size of the partial mixing zone (e.g, see Table~\ref{tab:c13}).

\begin{table*}
 \begin{center}
  \caption{$s$-process indicators from each stellar model. These are calculated from the average composition of the wind ejected
  over the star's life.}
 \label{tab:sinds}
  \vspace{1mm}
   \begin{tabular}{cccccccc}
    \hline\hline
Mass/$\Msun$ & Mdot & $M_{\rm mix}/\Msun$ &  [Rb/Zr] & [ls/Fe] & [hs/Fe] & [hs/ls] & [Pb/hs] \\ \hline
1.00 & VW93 & 0.0   & $-$0.014 &  0.001 &  0.001 & -0.002 &  0.001 \\
1.15 & VW93 & 0.001 & $-$0.233 &  0.287 &  0.820 &  0.533 &  0.699 \\
1.15 & VW93 & 0.002 & $-$0.420 &  0.504 &  1.031 &  0.528 &  0.749 \\
1.15 & VW93 & 0.006 & $-$0.652 &  0.828 &  1.457 &  0.629 &  0.618 \\
1.25 & VW93 & 0.002 & $-$0.504 &  0.643 &  1.126 &  0.484 &  0.782 \\
1.25 & VW93 & 0.006 & $-$0.719 &  0.983 &  1.620 &  0.638 &  0.637 \\
1.50 & VW93 & 0.001 & $-$0.554 &  0.639 &  1.300 &  0.661 &  0.659 \\
1.50 & VW93 & 0.002 & $-$0.664 &  0.897 &  1.535 &  0.637 &  0.664 \\
1.50 & VW93 & 0.006 & $-$0.821 &  1.294 &  1.924 &  0.630 &  0.508 \\
1.75 & VW93 & 0.006 & $-$0.787 &  1.175 &  1.832 &  0.656 &  0.633 \\
2.00 & VW93 & 0.001 & $-$0.754 &  1.098 &  1.834 &  0.736 &  0.503 \\
2.00 & VW93 & 0.002 & $-$0.732 &  1.352 &  1.994 &  0.642 &  0.666 \\ 
2.00 & VW93 & 0.006 & $-$0.771 &  1.776 &  2.366 &  0.590 &  0.352  \\
2.25 & VW93 & 0.002 & $-$0.777 &  1.498 &  2.130 &  0.632 &  0.530 \\
2.50 & VW93 & 0.002 & $-$0.656 &  1.591 &  2.262 &  0.671 &  0.326 \\
2.50 & VW93 & 0.004 & $-$0.523 &  1.816 &  2.439 &  0.623 &  0.279 \\
2.75 & VW93 & 0.002 & $-$0.398 &  1.592 &  2.251 &  0.660 &  0.278 \\
3.00 & VW93 & 0.001 & $-$0.303 &  1.286 &  1.955 &  0.669 &  0.408 \\
3.00 & VW93 & 0.002 & $-$0.141 &  1.478 &  2.128 &  0.650 &  0.439 \\
3.25 & VW93 & 0.001 & $-$0.111 &  1.260 &  1.866 &  0.606 &  0.489 \\
3.50 & VW93 & 0.001 & $-$0.070 &  1.265 &  1.859 &  0.594 &  0.551 \\
3.75 & VW93 & $1\times 10^{-4}$ & $-$0.062 &  1.242 &  1.854 &  0.611 &  0.526 \\
3.75 & VW93 & 0.001 & $-$0.062 &  1.242 &  1.854 &  0.611 &  0.526 \\
3.75 & B95  & 0.001 & $-$0.079 &  1.178 &  1.784 &  0.606 &  0.532 \\
4.00 & B95  & $1\times 10^{-4}$ & 0.008 &  0.462 &  0.891 &  0.429 &  0.762 \\
4.00 & B95  & 0.001 & $-$0.063 & 1.204 &  1.801 &  0.597 &  0.584 \\
4.50 & B95  & 0.0   &  0.275 & 0.120 &  0.016 & $-$0.104 & $-$0.004 \\
4.50 & B95  & $1\times 10^{-4}$ & $-$0.023 &  0.471 &  0.900 & 0.429 & 0.807 \\
5.00 & VW93 & 0.0 & 0.540 &  0.852  & 0.205 & $-$0.647 & $-$0.149 \\
5.00 & B95  & 0.0 & 0.240 &  0.106  & 0.013 & $-$0.093 & $-$0.004 \\
5.50 & B95  & 0.0 & 0.182 &  0.071  & 0.008 & $-$0.062 & $-$0.002 \\
6.00 & B95  & 0.0 & 0.148 &  0.054  & 0.006 & $-$0.049 & $-$0.001 \\
6.50 & B95  & 0.0 & 0.131 &  0.046  & 0.004 & $-$0.042 &  0.000 \\
7.00 & VW93 & 0.0 & 0.166 &  0.065  & 0.004 & $-$0.061 & $-$0.001 \\
7.00 & B95  & 0.0 & 0.099 &  0.033  & 0.002 & $-$0.031 &  0.000 \\
\hline\hline
  \end{tabular} 
 \end{center}
\end{table*}

The effect of a partial mixing zone and the \iso{13}C($\alpha$,n)\iso{16}O neutron source on the operation of the $s$-process is particularly noticeable for the 4.5$\Msun$ model. This model has the minimum mass with efficient HBB, with more nitrogen production compared to carbon, and is the maximum mass where we include a \iso{13}C pocket. Models with no \iso{13}C pocket only release neutrons from the \iso{22}Ne($\alpha$,n)\iso{25}Mg reaction operating during thermal pulses. The \iso{22}Ne source results in more Rb than Zr, evidenced by the [Rb/Zr] ratio of 0.28 and little heavy $s$-elements (e.g., Ba) or Pb. In contrast, the model with a \iso{13}C pocket show a strong over-production of $s$-process elements, with a final [Ba/Fe] $\approx 1$ and [Pb/Fe] $=1.8$. The ratio of [Rb/Zr] $\approx 0.0$ shows a mix of \iso{22}Ne and \iso{13}C neutron sources, unlike the lower mass models which have [Rb/Zr] ratios that are strongly negative.

\subsection{The effect of mass-loss on the nucleosynthesis}

\begin{figure*}
    \centering 
\includegraphics[width=0.95\columnwidth]{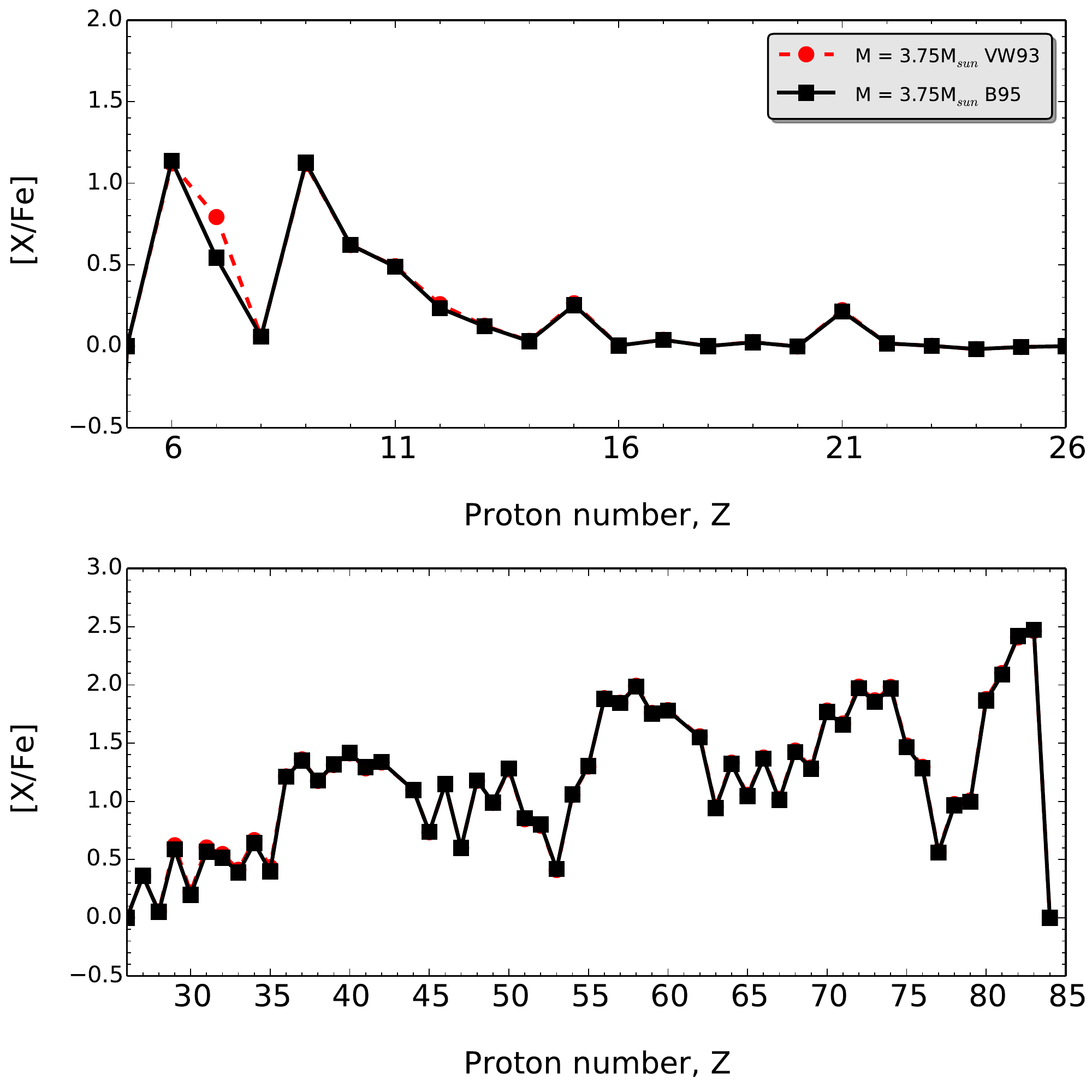}
  \caption{The final surface abundance predictions from the 3.75$\Msun$ models with B95 mass-loss (black filled squares) and VW93 mass-loss (red filled circles). Both models have $M_{\rm mix} = 0.001\Msun$.}
   \label{fig:m3.75compare}
\end{figure*}

We perform nucleosynthesis calculations for the 3.75$\Msun$, 5$\Msun$ and 7$\Msun$ models on the evolutionary sequences that use \citet[][hereafter VW93]{vw93} and \citet[][hereafter B95]{bloecker95} mass-loss on the AGB. The 3.75$\Msun$ model show mild HBB but also strong third dredge-up (e.g., in Fig.~\ref{fig:mdredge}) compared to the other intermediate-mass models. This model becomes strongly enriched in carbon and $s$-process elements regardless of the number of thermal pulses. Indeed, the VW93 model only experiences an extra 5~TPs compared to the B95 model. The resulting nucleosynthesis and yields are therefore similar as shown in Fig.~\ref{fig:m3.75compare}.

\begin{figure*}
    \centering 
\includegraphics[width=0.95\columnwidth]{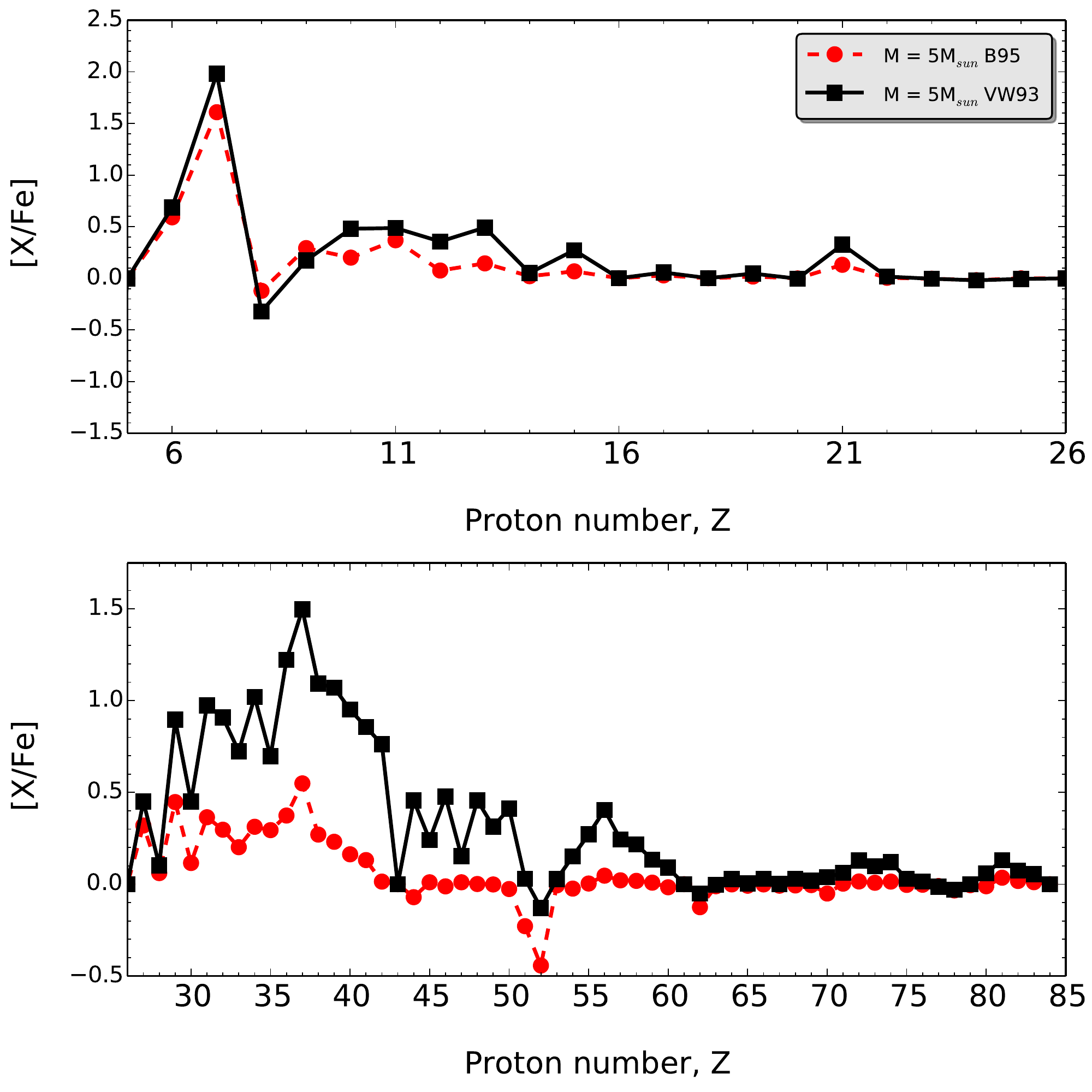}
  \caption{The final surface abundance predictions from the 5$\Msun$ models with B95 mass-loss (black filled squares) and VW93 mass-loss (red filled circles). Predictions for light elements are shown in the top panel and for heavy elements in the bottom panel.}
   \label{fig:m5compare}
\end{figure*}

The 5$\Msun$ and 7$\Msun$ models with B95 mass-loss show strong HBB on the evolution of the light elements and little increase in the heavy elements produced by the $s$-process. This is because the number of thermal pulses are considerably lower, as shown in Fig.~\ref{fig:tps}. In contrast, in Fig.~\ref{fig:m5compare} shows that the production of heavy-elements increases dramatically in the 5$\Msun$ model with VW93 mass-loss. In particular, the final surface [Rb/Fe] $\approx 1.5$ in the VW93 model compared to 0.43, an increase of an order of magnitude. Similar results, although less dramatic, are also observed for the 7$\Msun$ model where the final [Rb/Fe] $\approx 0.5$ in the VW93 model compared to [Rb/Fe] $=0.3$ in the model with B95 mass-loss (e.g., see results in Table~\ref{tab:sinds}).

Note that the 5$\Msun$ and 7$\Msun$ models become C-rich once the superwind begins, however as shown in Fig.~\ref{fig:co} for most ($\gtrsim 94$\%) of the thermally-pulsing AGB phase the surface composition is oxygen rich where C/O $< 1$. This result holds regardless of the AGB mass-loss rate used in the calculation.

In summary, the choice of AGB mass-loss rate is most important for intermediate-mass AGB models with strong HBB. A stronger mass-loss rate such as B95 favours stronger HBB on the surface composition and yields, while VW93 favours a strong overproduction of heavy elements formed by the $s$-process along with considerable primary nitrogen.

\section{Discussion} \label{sec:discuss} 

In this section we discuss observations of AGB stars and their progeny that are found with a similar metallicity to the AGB models presented here. We also discuss the issue of the light-element abundances observed in Galactic Globular Clusters, in comparison to our intermediate-mass model results. We start with barium stars, which are the product of mass transfer from an AGB companion. 

\subsection{Barium stars} 

Models of AGB stars at [Fe/H] $\approx -$0.7 are necessary to interpret the composition of Barium (Ba) stars because these stars are \textit{s}-process element enhanced first giant branch stars and dwarfs with spectral classes from G to K in the metallicity range [Fe/H] $\approx$ 0.3 to $-$1.

\citet{bidelman51} identified first the strong spectral features at specific wavelengths of elements heavier than iron (e.g. BaII at 4554 \AA, SrII at 4077 \AA, CH, CN, C$_{2}$ molecular bands) in the spectra of these stars. Later \citet{mcclure80} and \citet{mcclure83} pointed out the binary nature of Ba stars -- radial velocity observations showed that about 85\% of these stars belong to a binary system. Based on the evolutionary phase of the Ba stars and considering that \textit{s}-process elements are synthesized in the interior of AGB stars, the overabundance of s-elements in Ba stars cannot be intrinsic. \citet{mcclure84} suggested that the primary star -- now a white dwarf -- during the AGB phase transferred \textit{s}-process enhanced material to the secondary, which is now observed as a Ba star.  Because the temperature range of Ba stars is higher than that of AGB stars, their spectra are easier to model, which allows a more straightforward derivation of the abundances. Consequently Ba stars are an ideal laboratory to test AGB \textit{s}-process models.

Models at metallicity [Fe/H] $\approx -$0.7 are crucial to compare to the observations because within the current scenario of the \iso{13}C neutron source we expect the maximum production of the second peak \textit{s}-process elements at this metallicity. This is because this neutron source is primary, meaning that the number of free neutrons increases with decreasing Z. For higher metallicities the first peak is produced and for lower metallicities the third peak at lead is produced \citep[see e.g.,][]{gallino98,busso01}.
With the models presented here, together with the earlier published yields at different metallicities (\citealp{karakas16} and \citealp{fishlock14b}) we are now able to cover the whole range of the observed metallicities for Ba stars. 
Work is in progress (Cseh et al., in prep.) on a detailed comparison to the largest self-consistent sample of high resolution spectroscopic observations of Ba stars \citep{decastro16} as well as to the set of AGB predictions from the FRUITY database \citep[see][]{cristallo09,cristallo11,piersanti13,cristallo15}.

\subsection{AGB stars} 

Optical spectroscopic surveys of visually bright AGB stars in the Small
Magellanic Cloud show a lack of high luminosity (brighter than M$_{\rm bol}$$\sim$$-$6) C-rich AGB stars \citep{smith89,smith90b,plez93,smith95}.
The more luminous SMC-AGB stars ($-$7 $\leq$ M$_{\rm bol}$  $\leq$ $-$6) are O-rich and display s-process overabundances (as suggested by the strong ZrO and La II spectral features present in their optical spectra); $\sim$80\% of them are Lithium-rich with
log$\varepsilon$(Li)$\sim$1.0$-$3.5 \citep[e.g.,][]{smith95}. Their Li and s-process enrichments confirm that these stars are experiencing HBB and third dredge-up; i.e., HBB-AGB stars. In a sample of five HBB-AGB stars in the SMC, \citet{smith89} used plane-parallel model atmospheres and found average values of [Y/Fe]=$+$0.76$\pm$0.30 and [Zr/Fe]=$+$0.59$\pm$0.24, which are similar to those found in higher metallicity Galactic S-type AGB stars. A more detailed/reliable chemical abundance analysis (using spherical model atmospheres more appropriate for giant stars and extending the spectral coverage to the near-IR) of seven SMC HBB-AGB stars was carried out by \citet{plez93}.  They found an average metallicity of [z/H]=$-$0.52$\pm$0.13 and their study included for the first time other s-process elements like Rb and Nd as well as the $^{12}$C/$^{13}$C ratio. The Li-rich HBB-AGB stars in the SMC are C-poor ([$^{12}$C/H]$\approx$-$1.0$, but quite uncertain) and display very low C isotopic ratios ($^{12}$C/$^{13}$C=6.5$\pm$1.9), as expected from HBB models. However, these stars are not enriched in Rb ([Rb/z]$\leq-0.90\pm0.25$) but rich in other s-process elements such as Zr ([Zr/z]=$+$0.12$\pm$0.25) and Nd ([Nd/z]=$+$0.70$\pm$0.21). The unexpectedly low Rb abundances coupled with the Zr and Nd enhancements, led these authors to suggest that these low-metallicity HBB-AGB stars produce s-process elements at low neutron density via the $^{13}$C neutron source \citep[see also][]{abia01}. 

In strong contrast with the visually bright SMC HBB-AGB stars, higher metallicity Li-rich HBB-AGB (OH/IR) stars in our Galaxy are more obscured by dust, Rb-rich and Zr-poor, and show very high [Rb/Zr] ratios more typical of high neutron density and the $^{22}$Ne neutron source \citep{garcia06,garcia07a}. \citet{garcia09} carried out a high-resolution optical spectroscopic survey of the most obscured (and luminous) O-rich stars (including most of the known OH/IR stars) in the Magellanic Clouds (MC) and uncovered the low-metallicity Rb-rich AGB counterparts (both in the LMC and SMC); because of HBB flux excess, the MC Rb-rich AGB stars are even brighter than M$_{\rm bol}$$\sim$$-$7. \citet{garcia09} suggest that they might have progenitor masses of at least $\sim 6-7 \Msun$, while the visually bright Rb-poor HBB-AGB stars should have lower masses (say $\sim 4-4.5\Msun$). 

The large uncertainty affecting the Rb abundance and the
possibility of non-LTE effects means a direct comparison with model predictions is not straight forward. In addition, the Rb I lines can be strongly affected by circumstellar effects (especially in the obscured Rb-rich AGB stars) and exploratory pseudo-dynamical models show that the derived Rb abundances might be much lower (by 1$-$2 dex) when using extended model atmospheres \citep{zamora14,perez17}.
Thus a quantitative comparison with the model predictions should be restricted to the Li, Zr, and Nd abundances and $^{12}$C/$^{13}$C ratios observed in the visually bright SMC HBB-AGB stars by \citet{plez93}, while a qualitative comparison may still be applicable to Rb and the most extreme and obscured (OH/IR) Rb-rich stars.

Most of the intermediate-mass AGB models with HBB we have calculated here do not produce a strong $s$-process enrichment at the surface. These models would not be able to match the abundances measured by \citet{plez93}. The models with a strong enhancement in $s$-process abundances are the 4.5$\Msun$ model with a \iso{13}C pocket and the 5$\Msun$ model with \citet{vw93} mass-loss on the AGB (and to a lesser extent the 7$\Msun$ with VW93 mass-loss). In the \citet{plez93} sample the average [Zr/z] $\approx +0.25$, which is lower than the final [Zr/Fe] predicted at the end of the TP-AGB in both the 4.5$\Msun$ and 5$\Msun$ models ([Zr/Fe] $\gtrsim 0.6$). We are likely observing the real AGB stars some time during their TP-AGB phase, not at the end. Examining the 5$\Msun$ in more detail, at thermal pulse number \#55, the [Zr/Fe] $\approx 0.25$, C/O $\approx 0.5$, \iso{12}C/\iso{13}C $\approx 3$, which matches the observations of the Plez stars quite well. The predicted [Nd/Fe] $\approx -0.01$ at thermal pulse \#55 is much lower than the average of 0.70. Note that the final [Nd/Fe] = 0.1 in the 5$\Msun$ model, which is consistent with expectations from the \iso{22}Ne source but at odds with the observations. Furthermore, the Li abundance at pulse \#55 is much lower than observed (see above), where $\epsilon$(Li) $= -1$, down from the peak value of $\log \epsilon$(Li) = 4.25 at pulse \#10.

The high abundance of Nd points to a contribution from the \iso{13}C neutron source, so we examine the 4.5$\Msun$ model with a \iso{13}C pocket. At the 16th thermal pulse [Zr/Fe] $\approx 0.25$ and [Nd/Fe] $=0.5$, which is much more promising compared to observations. However, the C/O ratio $= 0.94$ at this stage although \iso{12}C/\iso{13}C $\approx 3$, which indicates that HBB is just getting going. By the 20th thermal pulse, C/O ratio has dropped to $= 0.20$, while the $s$-process elements are still increasing, to [Zr/Fe] = 0.33 and [Nd/Fe] = 0.61, respectively. We also note that the model star is enhanced in Li at this stage, with $\log \epsilon$(Li) = 3.2, which is in better agreement with the observations mentioned above.

We conclude that $s$-process rich, intermediate-mass stars observed by \citet{plez93} are of lower mass than suggested in the paper, and are closer to 4.5$\Msun$ than 5 or 7$\Msun$, and show evidence of the \iso{13}C neutron source \citep[see also][]{garcia06,garcia09}. While we are unable to make a definitive conclusion as to the best AGB mass-loss rate the predictions for Li and the $s$-process elements point toward the \citet{bloecker95} mass-loss rate. Indeed, the best fit model uses the \citet{bloecker95} mass-loss rate, although the models with \citet{vw93} result in stronger overabundances of $s$-process elements for models with HBB.

\subsection{Post-AGB stars} 

We now focus on objects that have evolved beyond the AGB and limit our discussion to post-AGB stars, the progeny of AGB stars. While there are many studies of planetary nebulae in the SMC, the observations are generally limited to elements lighter than iron and we refer to studies by \citet{ventura16b} and \citet{garcia18}.

During the post-AGB phase, the warm stellar photosphere makes it possible to quantify photospheric abundances for a very wide range of elements from CNO up to some of the heaviest $s$-process elements well beyond the Ba peak 
\citep{vanwinckel03,reyniers03} that are brought to the stellar surface during the AGB phase. This is not possible with AGB stars since molecular veiling dominates their spectra \citep{abia08}. Therefore post-AGB objects 
and in particular the single stars, can provide direct and stringent constraints on the parameters governing stellar evolution and nucleosynthesis, especially during the chemically-rich AGB phase. 

The well constrained distances to the post-AGB stars in the 
LMC and SMC offer unprecedented tests for AGB theoretical structure
and enrichment models of single low- and intermediate-mass stars. However, 
owing to their short lifetime these objects are rare. So far, the photospheric 
chemistry of only a few single post-AGB stars in the LMC and SMC have been 
studied in detail \citep{desmedt12,vanaarle13,desmedt15,kamath17}.
The study by \citet{desmedt12} revealed the most $s$-process enriched
post-AGB star, J004441.04-732136.4, in the SMC. This object is a
single, luminous (L/\Lsun\,$\approx\,$7000) post-AGB object with a
low metallicity ($\textrm{[Fe/H]}$\,=\,$-$1.34$\pm$0.32) compared to the mean
metallicity of the young stars in the SMC. The estimated photospheric
C/O ratio (C/O\,=\,1.9\,$\pm$0.7) indicates that it is a 
C-rich source while the high $s$-process overabundances (e.g.,
$\textrm{[Y/Fe]}$\,=\,2.15, $\textrm{[La/Fe]}$\,=\,2.84) show that this star is extremely $s$-process enriched. Furthermore, J004441.04-732136.4 also shows the 
presence of the 21\,micron feature in its mid-infrared 
(mid-IR) spectra \citep{volk11}. The luminosity and chemistry of J004441.04-732136.4 point toward a star of initial mass $\approx 1.3\Msun$ that is self-enriched in carbon and $s$-process elements.

The most unusual features of J004441.04-732136.4 from a nucleosynthesis
point of view is its Pb abundance and C/O ratio. Though the theoretical stellar
models presented in \citet{desmedt12} predict an $s$-process distribution 
very similar to the observed one, the predicted Pb abundance is significantly higher than the observed Pb upper limit. Furthermore, while the predicted C
overabundance is compatible with the observations, the predicted O
abundance is significantly lower resulting in a predicted C/O ratio of
$\sim$\,18\,$-$\,20, which is clearly too high. This indicates that the
star obtained large enrichments of heavy elements, while keeping a low
C/O ratio. No isotopic abundances could be estimated, but the models 
predicted a very high $^{12}$C/$^{13}$C ratio ($\sim$\,1800), which is 
not yet constrained by observations. 

Other post-AGB stars in the LMC have also been found to have low-Pb
abundances, relative to AGB model predictions \citep{vanaarle13,desmedt14,desmedt15}. 
A systematic study of 14 post-AGB stars in the Galaxy \citep{desmedt16} combined with the results for the  MC objects find that the Pb discrepancy seems to occur in stars with [Fe/H] $< -0.7$ 
but is not present in more metal-rich post-AGB stars with [Fe/H] $> -0.7$. 
This cut-off is the metallicity of the current models. From Fig.~\ref{fig:surf} and Table~\ref{tab:sinds} we can see that the high [Pb/Fe] abundances predicted for the low-mass AGB stars is consistent with standard calculations of the $s$-process \citep{busso01,cristallo15} of a metallicity around [Fe/H] $\approx -0.7$. 
If we take the 1.25$\Msun$ model as an example, the final C/O = 2.33, [Ba/Fe] = 1.1, and [Pb/Fe] = 1.9 when $M_{\rm mix} = 2\times 10^{-3}\Msun$ is adopted.
It is clear that our models produce high [Pb/Ba] in contrast to the observations
of J004441.04-732136.4 and the other low-metallicity post-AGB
stars. 

The cause of the low-Pb abundance in these post-AGB stars has
not been identified but stellar rotation and/or a non-standard neutron-capture
nucleosynthesis have been invoked \citep{lugaro15}. The high O abundance of J004441
is also a mystery and suggests a high O intershell abundance of $\approx 0.1$ by mass, consistent with models by \citet{pignatari16} which include convective-boundary mixing at the base of the He-flash driven convective pocket \citep[see also][]{herwig00}.

Another single post-AGB star in the SMC is J005252.87-722842.92, which was 
revealed by \citet{kamath17} to be chemically peculiar. Their detailed chemical
abundance analysis revealed that J005252 shows an intriguing photospheric 
composition with  no confirmed carbon-enhancement (upper limit of
$\textrm{[C/Fe]}$\,$<$\,0.50) nor any traces of $s$-process elements. 
They derived an oxygen abundance of $\textrm{[O/Fe]}$\,=\,0.29\,$\pm$\,0.1. 
An upper limit for the nitrogen abundance could not be determined since
there were no useful nitrogen lines within their existing spectral
coverage.
The derived stellar parameters (i.e., luminosity, metallicity) 
of J005252.87-722842.92 indicate it is a single post-AGB star with a luminosity of 
$8,200L_{\odot}$ and metallicity [Fe/H] = $-1.18$. The progenitor would have
been a star of $\approx 1.5-2\Msun$, which should become carbon and $s$-process
rich according to standard stellar evolutionary models such as those 
presented here \citep[see also models by][]{fishlock14b, ventura15,cristallo15}.
However, the observations are in contrast with these predictions. \citet{kamath17} 
concluded that J005252-722842.9 very likely reveals a new stellar 
evolutionary channel whereby a star evolves without any of the
chemical enrichments associated with third dredge-up episodes. 

The above studies show that single post-AGB stars are chemically diverse 
and a few significant discrepancies exist between the observed and 
predicted abundances, especially in the case of Pb, and possibly the $^{12}$C/$^{13}$C ratio. To fully understand the observed chemical 
diversity of post-AGB stars and its implications on AGB evolution and 
nucleosynthesis, detailed abundance studies of a larger population of 
post-AGB stars with well constrained distances that cover a spread in 
luminosities and metallicity is needed.

\subsection{Globular cluster abundances} \label{sec:gc}

Star-to-star abundance variations of the light elements Li, C, N, O, Na, Mg and Al have been observed in every well studied globular cluster (GC) and indeed the anti-correlation between O and Na is considered a definitive signature that differentiates globular clusters from open clusters and field stars \citep{carretta06,carretta09}. The origin of the light element anti-correlations is not known but various hypotheses have been proposed including pollution from rapidly rotating massive stars \citep{decressin07}, intermediate-mass AGB stars \citep[e.g.,][]{ventura09a}, massive binary stars \citep{demink09}, and supermassive stars \citep{den14}. Yields from these sources have been used in various chemical evolution and dynamical models in order to test these hypotheses, with various levels of success \citep{fenner04,dercole08,dercole10,dercole12,bekki17}. See also \citet{bastian18} for a recent review of formation scenarios.

One of the most discussed hypotheses is that a generation of low-metallicity intermediate-mass AGB stars polluted the clusters when they were forming, which can qualitatively explain the signature of hot hydrogen burning that has been observed \citep[see e.g.][]{ventura16c}. Coupled with the fact that AGB stars have slow winds that can be retained by the clusters and the ejecta does not lead to variations of the iron-peak elements is also in their favour.  Detailed AGB models have either failed \citep{herwig04b,karakas06a} or have had success in matching the some of the abundance trends observed in GCs but generally not all \citep{dorazi13a,ventura14,ventura16c,dellagli18}. AGB models provide such a variety of results, successful or not, because the nucleosynthesis is very much dependent on the efficiency of convection in AGB envelopes and on the AGB mass-loss rate \citep{ventura05a,ventura05b,karakas14dawes,dantona16}.

Here we discuss the impact of the AGB mass-loss rate on the predictions for O and Na for models of metallicity [Fe/H] $=-0.7$.  We limit our discussion to models $M\ge 4\Msun$ because these stars experience hot bottom burning and have short lifetimes ($\tau \lesssim 120$~Myr) such that they could pollute a forming globular cluster.

\begin{figure}
    \centering 
\includegraphics[width=0.95\columnwidth]{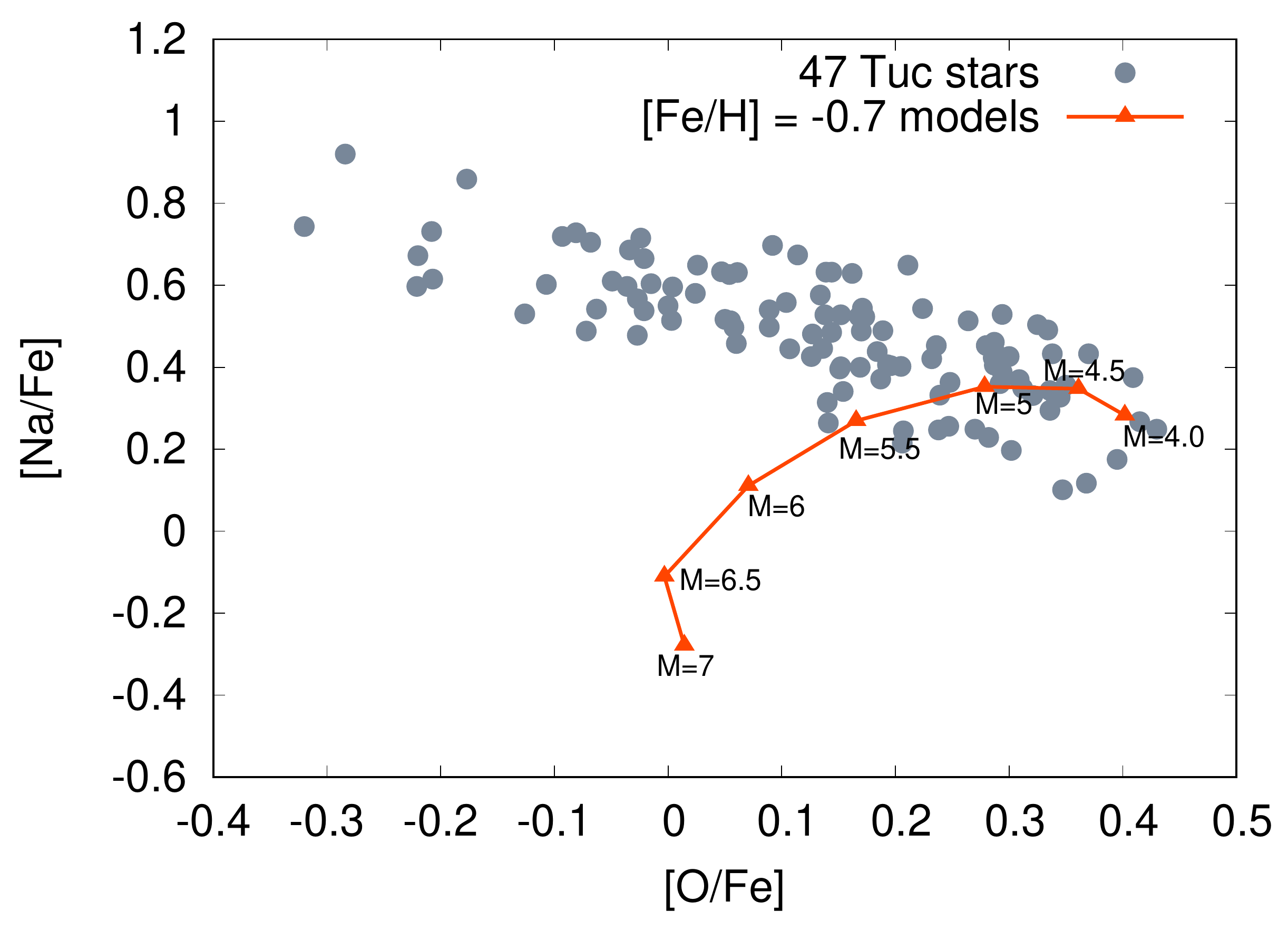}
  \caption{The [O/Fe] versus [Na/Fe] from the intermediate-mass models with HBB (filled orange triangles, with labels corresponding to the ejecta from each mass shown) compared to observed abundances in 47 Tucanae (filled grey circles) using data from \citet{carretta09}. Note that the models start with [O/Fe] = 0.0 so we have shifted the [O/Fe] abundances by $+0.4$~dex to reflect the primordial composition of metal-poor GCs which had [O/Fe] = $+0.4$.}
   \label{fig:o-na}
\end{figure}

Figs~\ref{fig:surf} and~\ref{fig:m5compare} show that the final surface composition of the 5$\Msun$ and 7$\Msun$ models are C and N rich, however the ejecta of the 7$\Msun$ is overall carbon-poor (for both the VW93 and B95 models). The ejecta of the 5$\Msun$ models are C-rich. If we examine the full mass range, models between 5.5$\Msun$ and 7$\Msun$ result in C-poor and N-rich ejecta, consistent with the observed C-N anti-correlation observed in GCs.  

In Fig.~\ref{fig:o-na} we show average [O/Fe] versus [Na/Fe] in the ejecta of the intermediate-mass models. We also compare our results to data from \citet{carretta09} for 47 Tucanae, which has the same mean metallicity of [Fe/H] $\approx -0.7$ as our models. While the data span a similar range in O abundance as the 47 Tuc stars the predicted trend between O and Na does not match the observations. This arises because in the most massive models, both O and Na are depleted together, a problem discussed previously by \citet{ventura08}.

We compare our theoretical predictions to the models by \citet{ventura16b}, who present models spanning a similar range in initial mass with a similar metallicity ($Z=0.002$).  The models by \citet{ventura16b} have the same mass-loss as the models shown in Fig.~\ref{fig:o-na} but more efficient convection owing to the Full Spectrum of Turbulence model to describe convection in stellar interiors (c.f. we use the Mixing Length Theory with $\alpha = 1.86$). 

We compare the range of O surface abundance predictions from our models compared to the $Z = 0.002$ models of \citet{ventura16b}. Using Fig.~1 from \citet{ventura16b}, we see that their 6$\Msun$ model ranges from $\log$ O/H $+12$ = 8 to 7, spanning an order of magnitude whereas in comparison our model of the same mass varies from $\log$ O/H $+12$ = 8 to 7.7. The 4.5$\Msun$ model from Ventura et al. also shows evidence of HBB, with a significant decline in O with a final $\log$ O/H $+12$ = 7.4. Our 4.5$\Msun$ in contrast shows no decline in O (see Fig.~\ref{fig:o-na}) owing to efficient TDU and less efficient HBB at this mass. 

In our models we have employed the \iso{22}Ne(p,$\gamma$)\iso{23}Na reaction rate from \citet{iliadis10}, however, recent experimental data from the Laboratory for Underground Nuclear Astrophysics have provided a re-evaluation of this rate up to 20 times higher \citep{cavanna15}. A full discussion on its impact on AGB models in relation to GC observations can be found in \citet{slemer17}.

\subsection{Mg isotopes} 

Magnesium isotopic observations are important because they can trace stellar and galactic evolution on timescales ranging from the very short to the long timescales involved with AGB stars. That is because the dominant isotope, \iso{24}Mg, is predominantly produced in short-lived core collapse supernova explosions while the neutron-rich isotopes \iso{25}Mg and \iso{26}Mg are produced in massive stars and in intermediate-mass AGB stars \citep[e.g.,][]{karakas03b}.

From the point of view of galaxy archeology theoretical yields are important to galactic chemical evolution models,  which when compared to observations can set characteristics of our galaxy. Several studies have investigated magnesium isotopic abundances in order to shed light on stellar and galactic chemical evolution 
\citep{barbuy85,barbuy87a,barbuy87b,yong03a,yong03b,yong04,gay00,melendez07}.

It is possible to find in the literature observations of stars with metallicities similar to [Fe/H] $\approx -0.7$, such as \citet{yong06b}, \citet{melendez09} and \citet{thygesen16}, for the M71 and 47 Tucanae globular clusters. The study of \citet{melendez09} of nine giant M71 stars shows two different populations, CN-weak (pristine) giants with low magnesium isotopic ratios (\iso{26}Mg/Mg $\sim4\%$) and CN-rich (polluted) giants with higher magnesium isotopic ratios (\iso{26}Mg/Mg $\sim 8\%$) suggest that a second generation of stars was polluted by a generation of intermediate-mass AGB stars. The recent analysis of Mg isotopes in 13 RGB stars by \citet{thygesen16}  examine the pollution mechanisms in order to explain the multiple population phenomenon present in the 47 Tuc. However, in contrast with the study of \citet{melendez09}, \citet{thygesen16} do not find evidence of different magnesium isotopic ratios for the pristine population in comparison with the polluted one. If we consider AGB stars as the main source of heavy Mg isotopes, the yields are therefore crucial in order to understand the multiple population scenario (e.g., see also discussion in \S\ref{sec:gc}).  

\begin{figure}
    \centering 
\includegraphics[width=0.95\columnwidth]{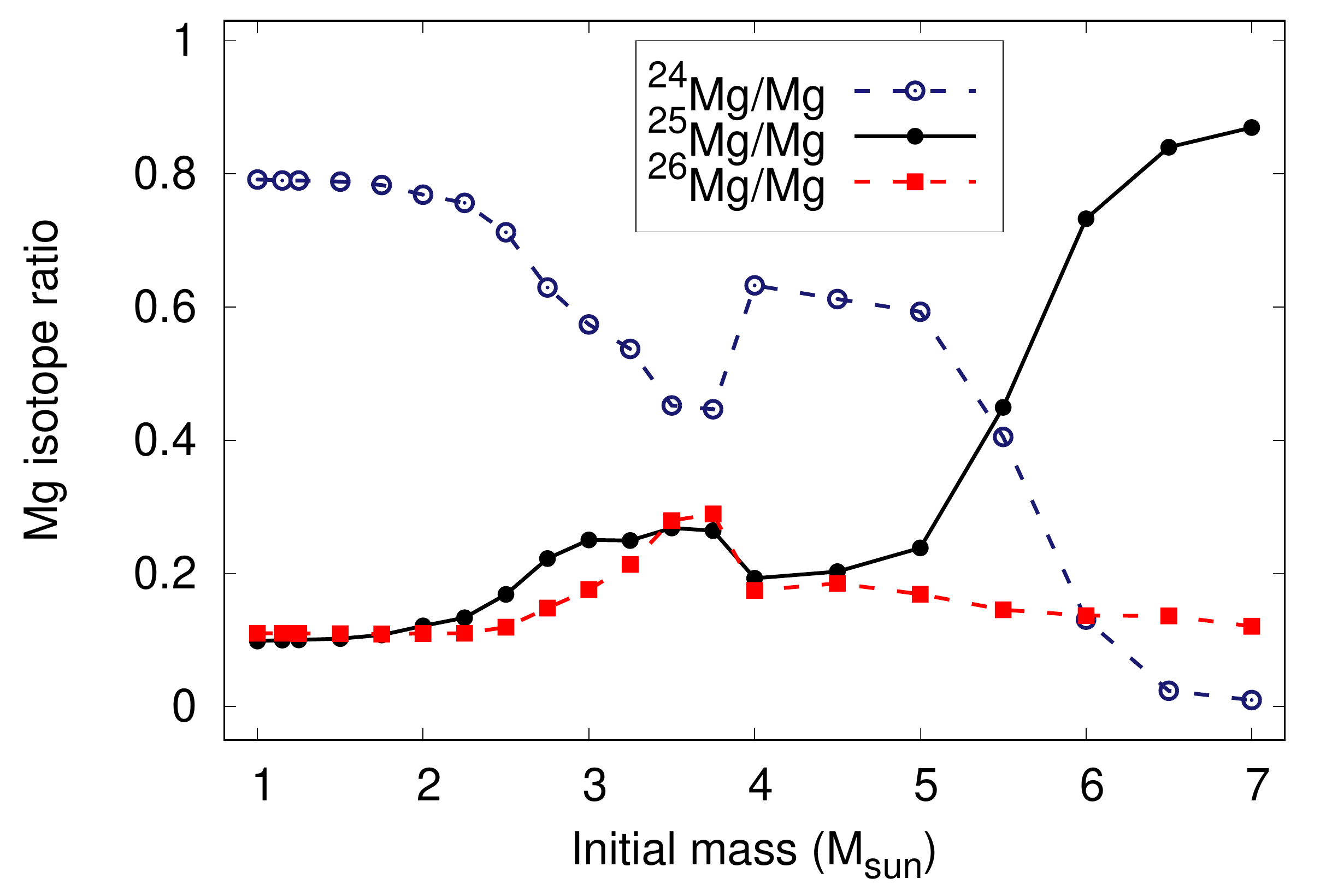}
  \caption{Ratios of \iso{24}Mg, \iso{25}Mg and \iso{26}Mg to the total Mg content at the surface of the star at the end of the calculation. The initial Mg isotope ratios are solar, where \iso{24}Mg/Mg = 0.789, \iso{25}Mg/Mg = 0.10 and \iso{26}Mg/Mg = 0.11.}
   \label{fig:mg}
\end{figure}

In Fig.~\ref{fig:mg} we show the Mg isotopic ratios relative to the total Mg content at the surface as a function of stellar mass for all of the $Z=0.0028$ models. This figure illustrates that low-mass AGB stars below $\approx 2\Msun$ do not result in observable variations to the Mg isotope ratios from solar. In contrast, models with third dredge-up and neutron-capture nucleosynthesis (but not HBB) range from 2-4$\Msun$ and show significant variations. Here the neutron-rich Mg isotopes are produced by \iso{24}Mg(n,$\gamma$)\iso{25}Mg(n,$\gamma$)\iso{26}Mg. Models between 4-5$\Msun$ experience mild HBB but do not dredge-up as much material from the He-shell as $\approx 3\Msun$ models, hence their neutron-rich Mg isotopic ratios are above solar but not significantly so. The most dramatic changes are for the intermediate-mass AGB models above 5$\Msun$ as a result of strong HBB. Here the variations are caused by proton-captures involving the Mg-Al chain reactions, including a strong depletion of \iso{24}Mg \citep[see also][]{izzard07}. Our predictions show strong enhancements in \iso{25}Mg and little variation to \iso{26}Mg, in contrast observations which show enhancements of \iso{26}Mg and no variation to \iso{25}Mg \citep[e.g.,][]{yong03a}.

\section{Conclusions} \label{sec:conclude}

In this study we present new theoretical models of low and intermediate-mass stars between 1$\Msun$ to 7$\Msun$ with a metallicity $Z=0.0028$ or [Fe/H] $\approx -0.7$. We present new evolutionary sequences, surface abundances (both elemental and isotopic), and stellar yields for all stable elements between C and Bi.

The stellar models have been calculated with updated input physics compared to the models in \citet{karakas10a}, which are widely used in studies of Galactic chemical evolution \citep[e.g.,][]{romano10,kobayashi11a}. For the AGB mass-loss rate, we continue to use the \citet{vw93} mass-loss rate for low-mass AGB stars that do not experience HBB, although for intermediate-mass stars we perform evolutionary calculations with both \citet{vw93} and the \citet{bloecker95} mass-loss rate, and adopt the latter for post-processing nucleosynthesis calculations. We calculate nucleosynthesis models for three intermediate-masses ($M=3.75, 5, 7\Msun$) using both mass-loss prescriptions. We find that the mass-loss rate does not play a strong role in the final surface abundances and yields for stars $\lesssim 4\Msun$ but can have a dramatic effect on stars around 5$\Msun$. 

We discuss some of the relevant observations that the new predictions can be compared with including AGB and post-AGB stars in the Small Magellanic Clouds, barium stars, and globular clusters. In particular, we discuss our models against the observations by \citet{plez93} and conclude that those stars are of lower mass than originally thought (closer to 4.5$\Msun$ than 5 or 7$\Msun$) and the stars show evidence of both the \iso{13}C and \iso{22}Ne neutron source in their surface compositions. While we are hesitant to make a definitive conclusion it does seem that the \citet{bloecker95} mass-loss rate produces AGB models that better match the bright O-rich stars in the \citet{plez93} sample.

Finally the new yields will also be useful for studies of specific populations of meteoritic stardust grains that originated in AGB stars (see Appendix A) and of Galactic chemical evolution, providing a missing link between the yields of solar metallicity presented in \citet{karakas16} and the lower metallicity yields presented by \citet{fishlock14b} and \citet{shingles15}. 

\section*{Acknowledgements}

The authors thank the referee for providing comments that have helped improve the discussion in the paper.
AIK acknowledges financial support from the Australian Research Council (DP170100521 and FT110100475), and thanks Kenji Bekki for interesting discussions about globular clusters.  ML is a Momentum (``Lend\"ulet-2014'' Programme) project leader of the Hungarian Academy of Sciences. This research was supported in part by the National Science Foundation under Grant No. PHY-1430152 (JINA Center for the Evolution of the Elements). MC would like to acknowledge support from the Capes/PDSE Grant (88881.135113/2016-01). BC was supported by a STSM Grant from COST Action CA16117. DAGH acknowledges support provided by the Spanish Ministry of Economy and Competitiveness (MINECO) under grants AYA--2014--58082--P and AYA--2017--88254--P.




\bibliographystyle{mnras}
\bibliography{mnemonic,library} 




\appendix

\section{Examples of the on-line only supplementary data tables} \label{app:datafiles}

In this section we provide examples of the data-files available for download. In Tables~\ref{tab:examplesurf} and~\ref{tab:exampleiso} we show an example of the elemental and isotopic surface abundances for the 3.5$\Msun$, $Z=0.0028$ model with $M_{\rm mix} = 1 \times 10^{-3}\Msun$. In Table~\ref{tab:exampleyield} we show an example of the stellar yields available for download for the 2$\Msun$, $Z = 0.0028$ model with $M_{\rm mix} = 2 \times 10^{-3}\Msun$.

\begin{table*}
 \begin{center}
  \caption{Example of the surface abundance tables available on-line. We show the first few lines at the
beginning of the 3.5$\Msun$, $Z = 0.0028$ model table, and the first few lines after thermal pulse \#10.}
 \label{tab:examplesurf}
  \vspace{1mm}
   \begin{tabular}{ccccccc}
    \hline\hline
\multicolumn{7}{l}{\# }\\
\multicolumn{7}{l}{\# Initial mass = 3.500, Z =  0.0028, Y =  0.250, $M_{\rm mix}$ = 1.00E-03}\\
\multicolumn{7}{l}{\# }\\
\multicolumn{7}{l}{\# Initial abundances} \\
\#El & $Z$  & $\log e(X)$ & [X/H] & [X/Fe] & [X/O] & $X(i)$ \\
\multicolumn{7}{l}{...}\\
\multicolumn{7}{l}{\# TP \hspace{0.2mm}  Mass   \hspace{3mm}   Mcore   \hspace{3mm}  Menv   \hspace{3mm}  log L} \\ 
\multicolumn{7}{l}{\# 103.499160 0.844551 2.654610 4.303790} \\
 \#El & $Z$  & $\log e(X)$ & [X/H] & [X/Fe] & [X/O] & $X(i)$ \\ 
  p  & 1 & 12.000000 &   0.022522  & 0.733065 & 0.732485 & 2.62931E-01 \\
  c  & 6 &  8.285793 & $-$0.184207 & 0.526336 & 0.525756 & 1.69944E-03 \\
  n  & 7 &  7.610781 & $-$0.259219 & 0.451324 & 0.450744 & 4.18853E-04 \\
  o  & 8 &  8.024012 & $-$0.705988 & 0.004555 & 0.000000 & 1.23891E-03 \\
  f  & 9 &  4.203442 & $-$0.256558 & 0.453985 & 0.449430 & 2.22374E-07 \\
...  \\
\multicolumn{7}{l}{\# Elemental abundance ratios:} \\
\multicolumn{7}{l}{\# He/H =  8.9644E-02, C/O =  1.8271E+00, N/O = 3.86162E-01} \\
\hline
\hline
  \end{tabular} 
\\
 \end{center}
\end{table*}

\begin{table*}
 \begin{center}
  \caption{Example of the isotopic abundance tables available on-line. We show the first five lines of the 
3.5$\Msun$, $Z = 0.0028$ model table for the first five isotopic ratios in the table.}
\label{tab:exampleiso}
  \vspace{1mm}
   \begin{tabular}{cccccc}
    \hline\hline
\multicolumn{6}{l}{\# }\\
\multicolumn{6}{l}{\# Initial mass = 3.500, Z =  0.0028, Y =  0.250, $M_{\rm mix}$ = 1.00E-03}\\
\multicolumn{6}{l}{\# }\\
\multicolumn{6}{l}{\#Initial isotopic abundance ratios:}\\
\# c12/c13 &  n14/n15 &  o16/o17 &  o16/o18 & mg24/mg25 & ... \\
 8.940E+01 & 4.476E+02 & 2.632E+03 & 4.988E+02 & 7.899E+00 & ... \\
\multicolumn{6}{l}{\# }\\
\multicolumn{6}{l}{\# During TP-AGB}\\
\multicolumn{6}{l}{\# }\\
  1.876E+01 & 2.874E+03 & 2.023E+02 & 7.599E+02 & 8.021E+00 ... \\
  1.889E+01 & 2.846E+03 & 2.033E+02 & 7.579E+02 & 8.020E+00 ... \\
  1.993E+01 & 2.845E+03 & 2.034E+02 & 7.577E+02 & 8.020E+00 ... \\
  2.532E+01 & 2.846E+03 & 2.039E+02 & 7.585E+02 & 8.014E+00 ... \\
  3.480E+01 & 2.867E+03 & 2.050E+02 & 7.613E+02 & 7.977E+00 ... \\
...  \\
\hline
\hline
  \end{tabular} 
\\
 \end{center}
\end{table*}

\begin{table*}
 \begin{center}
  \caption{Example of the yield tables available on-line.  We show the first five rows of the 2$\Msun$, $Z = 0.0028$ model yield table with $M_{\rm mix} = 2\times 10^{-3}\Msun$.}
 \label{tab:exampleyield}
  \vspace{1mm}
   \begin{tabular}{cccccccc}
    \hline\hline
\multicolumn{8}{l}{\# Initial mass = 2.000, Z =  0.0028, Y =  0.250, $M_{\rm mix}$ = 2.00E-03} \\
\multicolumn{8}{l}{\# Final mass = 0.659, Mass expelled =   1.3410} \\
 \#El & $Z$  & $\log e(X)$ & [X/H] & [X/Fe] & $X(i)$ & Net M($i$) & Mass($i$) \\
  p &  1 & 12.000000 & 0.000000 & 0.000000 & 7.06606E-01 & $-$6.24086E-02 & 9.47559E-01 \\
 he &  2 & 11.007096 & 0.077096 & 0.769786 & 2.85232E-01 & 4.70099E-02 & 3.82497E-01 \\
  c &  6 &  9.061618 & 0.591618 & 1.284307 & 9.70318E-03 & 1.23319E-02 & 1.30120E-02 \\
  n &  7 &  7.602188 & $-$0.267812 & 0.424878 & 3.92871E-04 & 3.27618E-04 & 5.26840E-04 \\
  o &  8 &  8.145969 & $-$0.584030 & 0.108660 & 1.56957E-03 & 4.78468E-04 & 2.10479E-03 \\
  f &  9 &  5.112539 & 0.652539 & 1.345229 & 1.72569E-06 & 2.20903E-06 & 2.31416E-06 \\
\multicolumn{8}{l}{...} \\
\hline\hline
  \end{tabular} 
 \end{center}
\end{table*}

\section{Relevance of low-metallicity AGB models to the interpretation of meteoritic stardust grains}

Predictions from AGB models of SMC metallicity are also essential for the interpretation of the composition and the origin of meteoritic stardust grains because this metallicity is currently believed the lower boundary for the parents stars from which they originated. The grains are recovered from primitive meteorites and represent tiny dust particles ($\mu$m and sub-$\mu$m sized) that formed around stars, novae, and supernovae. They kept their individuality as microcrystals from their stellar source site, throughout their residence time in the interstellar medium, during their incorporation in the proto-solar cloud, and inside meteorites \citep{zinner14}. As such, they give us direct information about the isotopic composition produced by nuclear reaction and mixing in their parent stars.

The vast majority of stardust grains originated in AGB stars, mostly of metallicity around solar \citep{hoppe97a,nittler97,lugaro03b,lugaro17}. However, peculiar
populations exist that appear to show the signature of C-rich AGB stars of
metallicity down to the SMC metallicity. The most famous example are silicon carbide (SiC) grains belonging to the so-called Z population
\citep{hoppe97b,zinner06}. These represent only roughly 1\% of the whole
stardust SiC inventory and are, on average, of smaller size then the mainstream grains ($\sim$90\% of all SiC grains), believed instead to have originated in C-rich AGB stars of metallicity around solar. Their isotopic composition can be analysed with very high precision and its interpretation gives us information on AGB nucleosynthesis at low
metallicity, independently complementing spectroscopic data. However,
the first step is to pinpoint the exact origin of the grains, in terms of the mass and metallicity of the parent stars. The Z grains show deficits in
\iso{29}Si and excess in \iso{30}Si, relative to \iso{28}Si and to the
solar composition. These are interpreted as the signature of AGB stars
of low metallicity because the low \iso{29}Si/\iso{28}Si ratio is a
signature of the initial composition of the star, where \iso{28}Si is
$\alpha$-enhanced relative to \iso{29}Si, a secondary isotope. On the
other hand, \iso{30}Si can be efficiently produced by neutron captures in
low-metallicity stars particularly via the \iso{32}S(n,$\gamma$)\iso{33}S(n,$\alpha$)\iso{30}Si chain, starting at the abundant \iso{32}S. One main issue is related to the \iso{13}C/\iso{12}C ratios measured in the
grains. These are lower than solar, in the same range, roughly 20 to 90, as
shown by the mainstream grains believed to have originated in AGB stars
of metallicity roughly solar. This is particularly puzzling since another small population of SiC, the Y grains (also 1\% only of all SiC) believed to come from AGB stars of metallicity in-between the mainstream and the Z grains show instead \iso{13}C/\iso{12}C ratios higher than solar \citep{amari01a}. One interpretation is that extra mixing processes in
low-mass AGB stars reduce the \iso{13}C/\iso{12}C ratio and that these processes are more efficient in the parent star of the Z grains \citep{hoppe97b,palmerini11}. On the other hand, via the interpretation the Si isotopic ratios on the basis of GCE and AGB models \citet{lewis13} concluded that type Z grains appear to have originated from AGB stars of higher mass, on average, than mainstream and Y grains. In this case, HBB could be responsible for the low \iso{13}C/\iso{12}C ratios. One way to disentangle this issue would be to compare model prediction for the elements heavier than Fe in AGB stars low-metallicity to the composition of these elements in SiC-Z. While such data is not available yet, it will become in the near future as technical progress is making analysis of stardust more efficient, for
example, using the new Chicago Instrument for Laser Ionization (CHILI)
instrument \citep{stephan16}. Comparison to model predictions such as
those reported here will allow us to better constraint the origin of
Z grains and the processes occurring in their parent stars.

Another type of stardust grain of which a large fraction appears to show the
signature of low metallicity AGB stars are high-density graphite grains
\citep{amari12}. This conclusion is mostly based on comparing the carbon and Kr isotopic compositions. It is expected that as the metallicity decreases, the
production of graphite in AGB stars is favoured with respect to that of SiC
\citep{sloan08}. Complementary analysis and interpretations of the
features and composition of SiC-Z and high-density graphite grains can
also help us shed light on the efficiency of dust formation for
different types of dust and how this varies with the stellar
metallicity. Future work will be dedicated to a detailed comparison
between our models and the composition of SiC-Z and high-density grains, and to provide detailed predictions for the isotopic ratios of elements heavier than Fe
that can be measured in the grains.

Finally, a minor fraction of oxide grains are rich in \iso{16}O with
respect to \iso{17,18}O and with respect to solar \citep[they are known
as the Group III grains][]{nittler97}. This composition has been taken
as the indication of an origin in low-mass low-metallicity AGB stars, due
to the $\alpha$-enhancement of \iso{16}O relatively to the secondary
isotopes \iso{17,18}O, as in the case of Si in SiC-Z. The low-mass
origin is required for the star to keep the signature of such initial
composition until the AGB phase when the dust forms, i.e., with no effect
of dredge-up episodes on the O isotopes. Also for these grains AGB
models of low-metallicity are required to confirm the current picture.


\bsp	
\label{lastpage}
\end{document}